\documentclass[twocolumn,twocolappendix]{aastex63}

\usepackage{amsmath}
\usepackage{threeparttable,booktabs}
\usepackage{amssymb}
\usepackage{enumitem}

\received{16 December 2020}
\revised{11 January 2021}
\accepted{21 January 2021}

\shorttitle{Empirical models of Jupiter}
\shortauthors{}

\graphicspath{{./}{Figures/}}

\def\lsim{\mathrel{\rlap{\lower 3pt \hbox{$\sim$}} \raise 2.0pt \hbox{$<$}}}
\def\gsim{\mathrel{\rlap{\lower 3pt \hbox{$\sim$}} \raise 2.0pt \hbox{$>$}}}

\newcommand{\comments}[1]{} %usage: \comments{}

\newcommand\T{\rule{0pt}{2.6ex}}       % Top strut
\newcommand\B{\rule[-1.2ex]{0pt}{0pt}} % Bottom strut
\newcommand{\Vector}[1]{\mathbf{#1}}
\newcommand{\sub}[1]{_{\text{#1}}}
\newcommand{\unit}[1]{\;\mathrm{#1}}

\newcommand{\metalt}{Movshovitz et al. (in prep.)}

\begin{document}

\title{Connecting gravity field, moment of inertia, and core properties in Jupiter \\ through empirical structure models}

\correspondingauthor{Benno A. Neuenschwander}
\email{benno.andreasneuenschwander@uzh.ch}

\author{Benno A. Neuenschwander}
\affiliation{Center for Theoretical Astrophysics and Cosmology, Institute for Computational Science, University of Zurich \\
Winterthurerstrasse 190, CH-8057 Z{\"u}rich, Switzerland}

\author{Ravit Helled}
\affiliation{Center for Theoretical Astrophysics and Cosmology, Institute for Computational Science, University of Zurich \\
Winterthurerstrasse 190, CH-8057 Z{\"u}rich, Switzerland}

\author{Naor Movshovitz}
\affiliation{Department of Astronomy and Astrophysics, University of California, Santa Cruz \\
1156 High St, Santa Cruz, California, USA
}

\author{Jonathan J. Fortney}
\affiliation{Department of Astronomy and Astrophysics, University of California, Santa Cruz \\
1156 High St, Santa Cruz, California, USA
}

\begin{abstract}

Constraining Jupiter's internal structure is crucial for understanding its formation and evolution history.
Recent interior models of Jupiter that fit Juno's measured gravitational field suggest an inhomogeneous interior and potentially the existence of a diluted core. 
These models, however, strongly depend on the model assumptions and the equations of state used. 
A complementary modelling approach is to use empirical structure models. 
These can later be used to reveal new insights on the planetary interior and be compared to standard models. 
Here we present empirical structure models of Jupiter where the density profile is constructed by piecewise-polytropic equations.
With these models we investigate the relation between the normalized moment of inertia (MoI) and the gravitational moments $J_2$ and $J_4$. 
Given that only the first few gravitational moments of Jupiter are measured with high precision, we show that an accurate and independent measurement of the MoI value could be used to further constrain Jupiter's interior. An independent measurement of the MoI  with an accuracy better than $\sim 0.1\%$ could constrain Jupiter's core region and density discontinuities in its envelope. 
We find that models with a density discontinuity at $\sim$ 1 Mbar, as would produce a presumed hydrogen-helium separation, correspond to a fuzzy core in Jupiter. 
We next test the appropriateness of using polytropes, by comparing them with empirical models based on polynomials. 
We conclude that both representations result in similar density profiles and ranges of values for quantities like core mass and MoI.

\end{abstract}


\section{Introduction} \label{sec:introduction}

Understanding the internal structure of Jupiter is a longstanding objective in planetary science and efforts in this direction go back decades \cite[e.g.,][]{Hubbard1968, Podolak1974, Decampli1979}. Such efforts are still ongoing and are of great importance because Jupiter's interior can provide clues on its origin and evolution \cite[e.g.,][]{2014prpl.conf..643H,HelledStevenson2017,Vazan2018,Muller2019}.
The main theoretical tools in this effort are structure models, designed to reproduce the measured planetary mass, radius, and gravitational field.

For Jupiter, the ongoing Juno mission has provided accurate measurements of its gravity field via radio tracking \citep{Folkner2017,NatureIess}.
These accurate gravity data further constrain internal models of Jupiter and therefore are used to determine Jupiter's bulk composition, as well as the distribution of the different chemical elements within the planetary interior \cite[e.g,][]{2017Wahl, Debras_2019}.
However, it should be kept in mind that the planetary composition and structure cannot be observed directly. Information about the composition and its depth dependence is inferred by fitting theoretical models to the available data \citep[see][and references therein]{Helled2018}.

The total potential $U(\Vector{r})$ of a planet in the rotating frame is given by the sum of the gravitational potential $V(\Vector{r})$ and centrifugal potential $Q(\Vector{r})$:
\begin{align}\label{eq:U_expansion}
    U(r, \theta) &= V(r, \theta) + Q(r, \theta) \nonumber \\
    &= -\frac{GM}{r} \left(1-\sum_{n=1}^{\infty}\left( \frac{a}{r}\right)^n J_n P_n(\cos\theta)\right) \\
    &\quad\text{ }+ \frac{1}{2}\omega^2r^2\sin^2(\theta) \nonumber,
\end{align}
where $r$ and $\theta$ are the distance and co-latitude, respectively, $G$ the gravitational constant, $M$ the planet's mass, $a$ the equatorial radius, $P_n(\cos\theta)$ the Legendre Polynomial of degree $n$, and $\omega$ the rotation rate. $J_n$ (also $J$-values) are the gravitational harmonics and are integrals of the planet's mass distribution $\rho(\Vector{r})$ over its volume. Their calculation requires knowledge of the planet's shape, itself determined by the potential. An iterative solution process converges to the self-consistent equilibrium shape and gravity. For a fluid planet in hydrostatic equilibrium, only the even order coefficients $J_{2n}$ are non-zero; dynamic effects as well as external perturbers (e.g. a large satellite) can give rise to non-zero odd coefficients and to additional terms not present in eq.~\eqref{eq:U_expansion}.

\begin{figure}
    \centering
    \includegraphics[width = 0.5\textwidth]{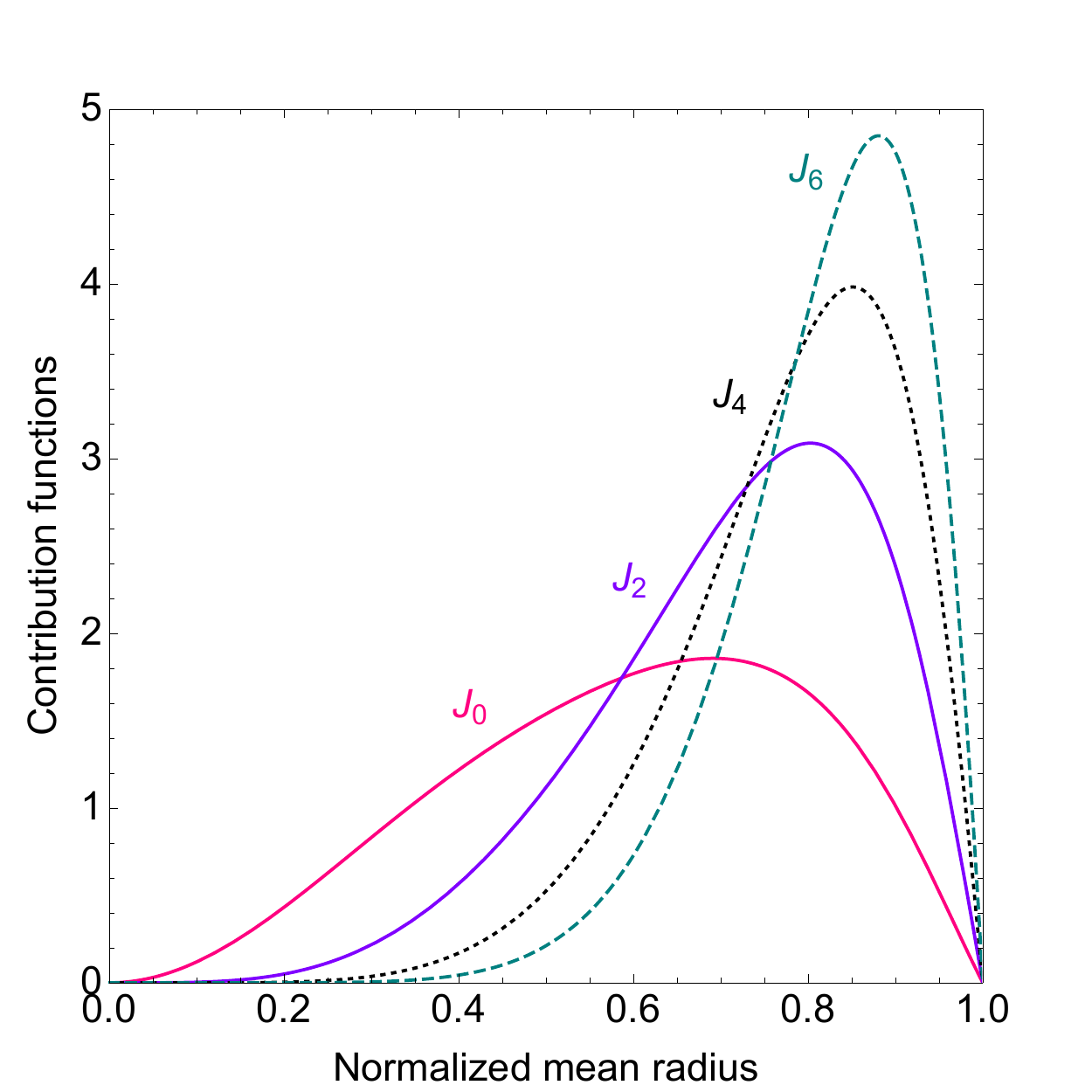}
    \caption{An example of calculated contribution functions of Jupiter (modified from  \citet{2011Helled}). One can calculate the $J$-value by integrating over the enclosed area (between the abscissa and the corresponding curve). However, for better visualization, each value is normalized. $J_0$ corresponds to the mass.}
    \label{fig:contr_function}
\end{figure}

Figure \ref{fig:contr_function} illustrates the contribution functions of the first four even $J$-values for Jupiter. The contribution functions are the normalized integrands of the gravitational moments and can be used to illustrate the ``weighting'' of various regions within the planet for a given $J_{2n}$ (e.g., \cite{ZHARKOV1974, Guillot2009}). It is clear from the figure that higher order coefficients are more sensitive to the outer regions of the planet and have a narrower and more pronounced region of sensitivity. 

Our understanding of Jupiter's interior has been challenged by Juno's measurements. 
Models reproducing the new data suggest that Jupiter's interior is inhomogeneous and display an extended core-envelope transition in the deep interior rather than a sharp boundary with a well-defined heavy-element core \citep{2017Wahl, Debras_2019}. 
These results challenge the simplified common view of giant planets being objects with a simple structure that can be separated into distinct layers.
In addition, these models imply that Jupiter's deep interior includes hydrogen and helium, and possibly composition gradients. This more complex internal structure must be explained by giant planet formation and evolution models \cite[e.g.,][]{HelledStevenson2017, Vazan2018, Muller2019}.   

It should be noted, however, that these recent internal structure models of Jupiter strongly depend on the equation of state (EoS) of the assumed composition; in particular the EoS of hydrogen under planetary conditions, the phase separation of helium \cite[e.g.,][]{Morales2013} and the behaviour of mixtures \cite[see][for recent review]{Helled2020}.
Necessarily, the planetary composition has to be \emph{assumed} by the modeler, of course with some free parameters in control. Even with robust formation theories and thermodynamic considerations directing these assumptions, there is still the risk of unavoidable uncertainties and biases ``contaminating'' the results.

While structure models that are based on physical EoSs generate detailed and easy to interpret models of Jupiter's composition and its depth dependence, there is also clear value in taking a complementary approach where the density profile is generated with a mathematical function, without direct reference to composition \cite[e.g.,][]{Helled2009,Helled2011_jup, Movshovitz2019, Ni2018}. A convenient approach is to use an empirical density profile based on polytropes \cite[e.g.,][]{Hubbard1975} or polynomials \citep[e.g.,][]{Helled2009,Movshovitz2019}.

Using gravitational data to describe and constrain a planet's interior yields non-unique solutions. In particular, it is hard to constrain the innermost region of a planet since the $J_{2n}$-values are ``blind'' to this part of the planet as shown in figure \ref{fig:contr_function}.

In this paper we address the following questions: (1) Is it possible to put some limits on Jupiter's core properties using only an accurate measurement of $J_2$ and $J_4$? (2) Is there  usable information in Jupiter's normalized moment of inertia (MoI) that is not degenerate with $J_2$ and $J_4$?
In order to answer these questions we construct a large range of empirical density profiles for Jupiter. In particular, we focus on the innermost region that can be viewed as representing a ``core''. We investigate the sensitivity of the calculated MoI, $J_2$, and $J_4$ values to the assumed core properties. 

Our paper is organized as follows. In section \ref{sec:methods} we explain the calculation method and the characteristics of our models. In section \ref{sec:results} we present and discuss the resulting density profiles. A summery and discussion are presented in \ref{sec:summary}.

\section{Methods} \label{sec:methods}
First, we generate density profiles of Jupiter that fit the measured gravitational coefficients $J_2$ \& $J_4$, as well as its mass, equatorial radius, and rotation period. Table \ref{tab:jupiter_properties} summarizes the planetary properties used for these models.  

Our empirical models are based on polytropes.
A polytrope describes the relation between the pressure $P$ and the density $\rho$ according to the free parameters $n$ and $K$: 
\begin{equation}
    P = K\rho^{1+\frac{1}{n}}.
    \label{eq:polytrope}
\end{equation}
Despite the simplicity of the function, it was found that polytropes can represent Jupiter's interior rather well \cite[e.g.,][]{Hubbard1975, Hubbard1999,Wisdom_2016}. \\ 
Although Jupiter's interior can be represented fairly well with a single polytrope, it is insufficient to fully fit its gravity data. In order to produce structure models that are consistent with Jupiter's gravity field and to explore a large parameter space we consider density profiles constructed with piecewise polytropes. That is, different polytropic relations hold in different radial regions of the interior. We allow up to three polytropes; up to three regions in the planet that have a different physical behavior.
The difference to traditional three-layer models is that the distinct regions, defined in our case by large differences in polytropic parameters, do not necessarily represent regions of homogeneous composition.
Solutions with consolidated polytropes, leading to fewer density jumps and fewer distinct regions are also permitted.

To facilitate description of the results we utilize the following notation. We designate the polytrope defining the outer region ${p_1}$, the one defining the middle region ${p_2}$, and the one defining the inner region ${p_3}$, each requiring two parameters, a coefficient $K_i$ and index $n_i$. Two additional model parameters define the transition radii between the different regions. $r_{1,2}$ is the radius where ${p_1}$ and ${p_2}$ meet, and $r_{2,3}$ is the radius where ${p_2}$ and ${p_3}$ meet, given as fractions of the planet's equatorial radius. The pressure $P$ and density $\rho$ at these special radii are sometimes of interest and are denoted with the same subscripts, e.g., $P_{1,2}=P(r_{1,2})$. 

We often think of the innermost region as representing Jupiter's core and refer to it as such, being careful to not assume it must be compact and/or composed primarily of heavy elements.
Depending on values of the parameters ${p_3}$ and $r_{2,3}$ this innermost region may instead represent a gradual increase in density (and therefore heavy elements), perhaps consistent with a diluted core. 
Note that several studies assume a constant density core, which is not physical for compressible material. An analysis of the validity of such a simplification is given in appendix \ref{subsection:CDCvsPC}.

A key question we aim to answer is whether the gravity field can be used to distinguish between a compact and a diluted core, and whether an independent measurement of the moment of inertia can help in this regard. We are therefore often interested in the radius and mass of this innermost region, and also designate them $r\sub{core}=r_{2,3}$ and $m\sub{core}$, respectively. 

Given a set of parameter values, we generate an interior density profile that, when in hydrostatic equilibrium, is consistent with the pressure implied by the polytropic relations. This is an iterative process. An initial guess for a density profile $\rho(r)$ is used to calculate the equilibrium shape and gravity, thereby implying a pressure profile $P(r)$ by hydrostatic equilibrium. The density is adjusted and the process repeated until $P(\rho(r))$ matches the polytropic relations everywhere in the planet. We keep the planet's mass, equatorial radius, and rotation period fixed.

The computationally time-consuming part of this process is the calculation of the equilibrium shape, a calculation that also yields the gravity coefficients $J_n$. We use an implementation of fourth-order Theory of Figures (ToF) \citep{ZharkovVladimirNaumovich1978Popi,Zharkov1970, Zharkov1975, Hubbard2014,Nettelmann_2017}, applicable to fluid planets in hydrostatic equilibrium with uniform rotation. Our calculation therefore neglects differential rotation or other dynamical effects. In reality, although hydrostatic equilibrium is expected to hold well in Jupiter's interior, there is evidence that observed surface winds penetrate to depth of $\sim$ 3000 km and influence Jupiter's gravity field \citep{Kaspi2018}. 
The zonal winds give rise to non-zero odd-numbered coefficients in eq.~\eqref{eq:U_expansion}, and also shift the even-numbered $J_{2n}$ relative to the values derived for static equilibrium \citep{Hubbard1982}. 
In principle, this offset could be calculated and accounted for, but this requires knowing the actual winds profile deep below the surface. Therefore we account for this offset by giving larger uncertainties to the measured $J$-values \citep[e.g.,][]{Kaspi2018,Guillot18}, compared with the formal measurement errors (see table~\ref{tab:jupiter_properties}).

ToF resolves the planet's shape on a finite set of equipotential levels. The more levels that are evaluated, the more precisely the planet's continuous interior is approximated. Our models employ 4096 levels, equally spaced in radius. The shape equations are evaluated explicitly on 128 equally spaced levels, and then spline-interpolated in the radial direction between them. This speeds up the calculation significantly while maintaining the desired precision. We validate this method by comparison with previously published results \citep{Militzer2019, Movshovitz2019}. An investigation of the impact of the model resolution (number of equipotential levels) on the calculated $J_{2n}$ and MoI is presented in appendix \ref{section:resoltution_dependence}.

We want to generate density profiles that exhibit a wide variety of core configurations, specifically, a wide range of $m\sub{core}$ and $r\sub{core}$ values. We therefore define a large discrete set of $(r\sub{core},m\sub{core})$ pairs in the range $0.025 \le{}r\sub{core}\le{}0.5$ and $1\le{}m\sub{core}\le{}100\unit{M_\oplus}$.
For each pair of values (core configuration) we run an unconstrained optimization algorithm to search for values of the model parameters that minimize our objective function $\mathcal{L}(J_2, J_4, m\sub{core})$: 
\begin{equation}
    \mathcal{L}(J_2, J_4, m\sub{core}) = A\cdot \delta J_2^2 + B\cdot \delta J_4^2 + C\cdot \delta m\sub{core}^2~,
\end{equation}
where
\begin{align}
   \delta J_2 &= \frac{J\sub{2,calc}-J\sub{2,obs}}{J\sub{2,obs}}, \\
   \delta J_4 &= \frac{J\sub{4,calc}-J\sub{4,obs}}{J\sub{4,obs}},\\ 
   \delta m\sub{core} &= \frac{m\sub{core,calc}-m\sub{core,conf}}{m\sub{core,conf}}.
\end{align}
$J\sub{2,obs}$ and $J\sub{4,obs}$ are the observed gravitational coefficients, $m\sub{core,conf}$ the core mass of the specific core configuration and $J\sub{2,calc}$, $J\sub{4,calc}$ and $m\sub{core,calc}$ the calculated model values. Changing the weights, $A$, $B$, and $C$, lets us nudge the optimization algorithm when it gets stuck in an unsuitable local minimum.

The search for model parameters is carried out by the simplex optimization algorithm \citep{Lagrias1998}\footnote{implemented in MATLAB's fminsearch}. If, for a certain core configuration, the algorithm fails to find values producing a model that fits $J_2$, $J_4$, and $m\sub{core}$ within their uncertainties or tolerance, resp., we conclude that the desired core configuration is invalid. We also invalidate some configurations based on central pressure and density. We exclude density profiles that result in central pressure greater than 100~Mbar \citep{Miguel2016,Debras_2019, 2017Wahl} or central density greater than $30,000\unit{kg\cdot m^{-3}}$, which is well above the expected density of rock at this pressure \citep{sesame7100, Musella2019,Thompson1974}.

\begin{threeparttable}
    \renewcommand\TPTminimum{\linewidth}
	\caption{Physical properties of Jupiter and its gravitational harmonics. $m\sub{rot}= \omega^2s^3/GM$ is the \textit{small parameter} used by ToF, where $\omega$ is the angular velocity, $s$ the mean radius, $G$ the gravitational constant and $M$ the planet's mass.}
	\begin{tabular}{l|ll}
			& \multicolumn{1}{c}{\textbf{Jupiter}} \\
		    \midrule
		Mass \tnote{1}		 	    & 317.8	     & [$\text{M}_\oplus$]\\  
		Equatorial radius \tnote{1}			& 71,492     & [\text{km}]\\ 
		Rotation period \tnote{2}	& 35,729.7	 & [$\text{s}$]\\ 
		$J_{2}$ \tnote{3}			& 14,696.572 & $[\times10^6]$\\ 
		$J_{4}$ \tnote{3}			& -586.609 & $[\times10^6]$\\ 
		$\Delta J_{2,\text{formal}}$ \tnote{3} & 0.014	 & $[\times10^6]$\\  
		$\Delta J_{2,\text{winds}}$ \tnote{4}  & 0.568  & $[\times10^6]$\\
		$\Delta J_{4,\text{formal}}$ \tnote{3} & 0.004	 & $[\times10^6]$\\  
		$\Delta J_{4,\text{winds}}$ \tnote{4}  & 0.2257  & $[\times10^6]$\\
		$m\sub{rot}$                   & 8.340783   & $[\times10^2]$\\
	\end{tabular} 
	\begin{tablenotes}
	    \item[1] \url{https://nssdc.gsfc.nasa.gov/planetary/factsheet/index.html} 
	    \item[2] \cite{RIDDLE1976}
	    \item[3] \cite{NatureIess}
	    \item[4] \cite{Kaspi2018}
	\end{tablenotes}
	\label{tab:jupiter_properties}
\end{threeparttable} \\

Note that the optimization algorithm returns a single local minimum. Therefore our models are clearly not the only possible three-polytrope representations of Jupiter, but are valid solutions. In future work we hope to use complementary algorithms to arrive a more complete description of the solution space. We also note that it is important to investigate in detail the impact of the model resolution (number of equipotential levels) on the inferred $J$-values and the MoI, as this can strongly affect the results. A preliminary analysis is presented in appendix \ref{section:resoltution_dependence} and we hope to address this more thoroughly in future research. 

\clearpage
\section{Results - empirical Jupiter models} \label{sec:results}
\begin{figure}
\centering
\includegraphics[width = 0.5\textwidth]{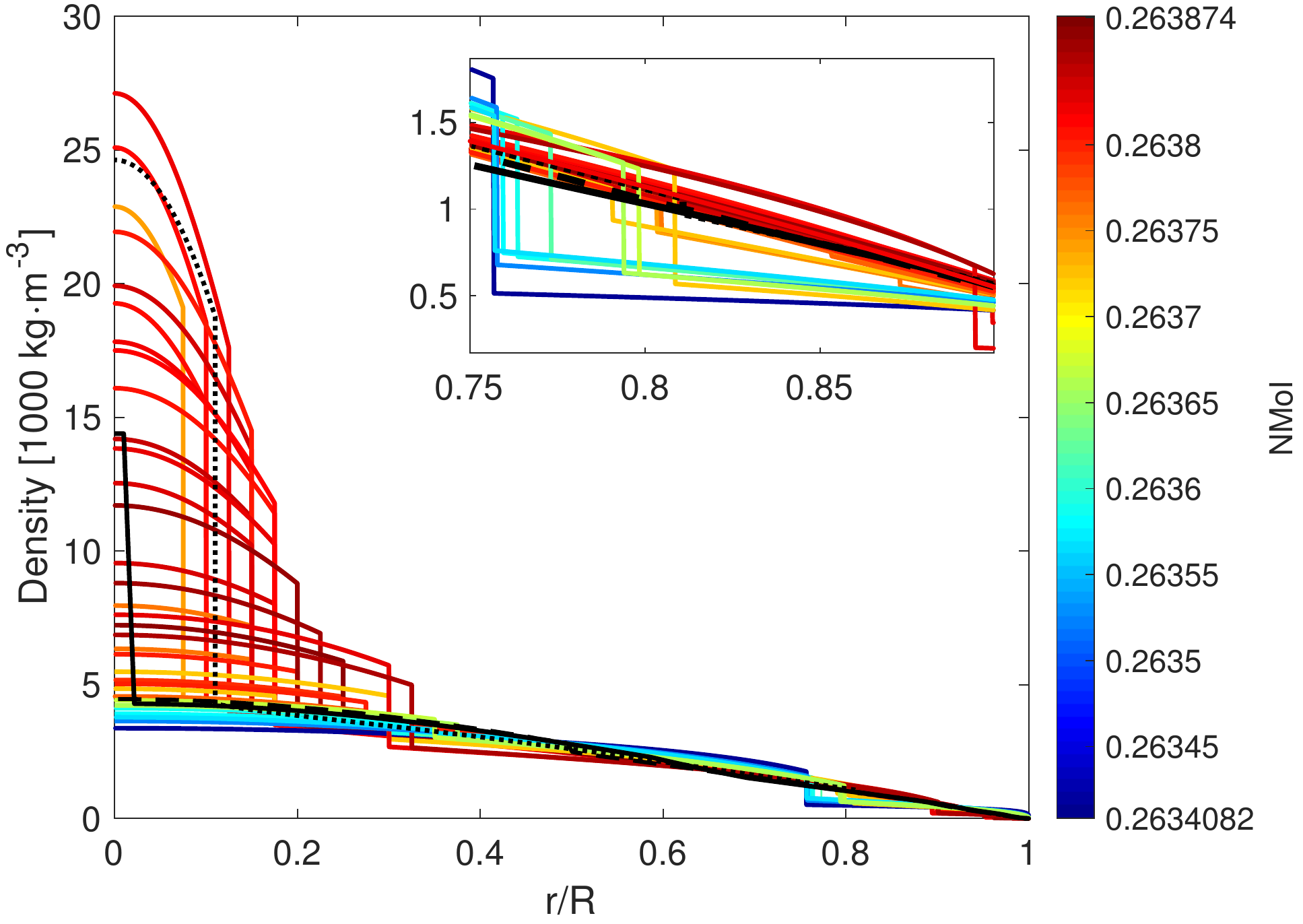}
\caption{Jupiter's density vs.~normalized radius of a representative selection of all \textit{good results}. The color of each solution illustrates its MoI value. For comparison, published results of \citet{Debras_2019} (black solid line), \citet{2017Wahl} (black dashed line) and \citet{Miguel2016} (black dotted line) are included. Most density discontinuities in the envelope, while not forced, tend to occur at $0.75 < r\sub{trans} < 0.9$. Diluted core solutions with low core densities tend to have larger density discontinuities at $r\sub{trans}$, resulting in a relatively low MoI value.}
\label{fig:all_in_one_MoI}
\end{figure}

Density profiles of Jupiter created with the procedure outlined above are presented in figure~\ref{fig:all_in_one_MoI}. Shown is a representative subset of all investigated density profiles that fit Jupiter's measured gravity field (\textit{good results}). The full solution space is shown in appendix \ref{sec:all_inferred_profiles}. The colors indicate the calculated MoI value. For comparison, three previously published composition-based models are overlaid. The solid black line is the model of \citet{Debras_2019} and the black dashed and dotted lines are solutions from \cite{2017Wahl} and \citet{Miguel2016}, respectively. 

Our piecewise-polytrope solutions space include solutions with lower central densities, potentially corresponding to a diluted core scenario, as well as solutions with sharp transitions to a central region of high density, implying compact and presumably rocky cores.
Independent of the various core properties, the density profile variation at a radial distance of $0.6 \le{r} \le{} 0.7$ is rather small. However, variations in the core region mostly affect the outermost region ($r \gtrsim 0.75$). 

Diluted cores with low core densities tend to have a larger density discontinuity at $r\sub{trans}$ which in turn results in a lower MoI value. This feature, together with an accurately measured MoI, can potentially be used to further constrain Jupiter's interior.

Interestingly, although we put no limits on the value of $r\sub{trans}$, we find that in most of the models where a large density jump occurs in the envelope, the transition radius is $\approx 0.75-0.9$. At these radii, densities around $\rho\sub{trans} \sim 250 -1500~\text{kg}\cdot \text{m}^{-3}$ and pressures around $P\sub{trans} \sim 0.5-3~\text{Mbar}$ occur.

A detailed analysis of the constraining power of the MoI with respect to $J_2$ and $J_4$ is presented in section~\ref{subsection:MoI_vs_J2} and \ref{subsection:MoI_vs_core}. While the connection between the MoI and the transition pressure (or radius) is presented in 
section~\ref{subsection:Transition_behavior}. A comparison to polynomial-based density profiles is shown in section~\ref{subsection:comparison_to_polynomials}.

\subsection{Relation between the gravitational moments and the MoI} \label{subsection:MoI_vs_J2}
The MoI and the second gravitational moment $J_2$ are closely correlated, both involving similar integrals over the density profile. The Radau-Darwin relation \citep[e.g.][]{Helled2011_jup} suggests that the two parameters are linked via the following relation:
\begin{equation}
    \text{MoI}= \frac{2}{3} \left[ 1 -  \frac{2}{5} \left( \frac{5m\sub{rot}}{m\sub{rot}+3J_2} -1 \right)^{1/2} \right],
    \label{eq:RadauDarwin}
\end{equation}
where $m\sub{rot}$ is the small parameter used in ToF (described in the caption of table \ref{tab:jupiter_properties}).
The Radau-Darwin relation is an approximation and it has been shown by several studies that there is no one-to-one correspondence between the MoI and  $J_{2}$.  
It may be that, at least in principle, knowledge of all gravity coefficients to high order and high precision would be enough to fully constrain $\rho(r)$ and therefore the MoI as well. But the more relevant question in practice is to what extent the MoI is already constrained by the measurable coefficients with their known uncertainties, and whether 
an \emph{independent} measurement of the MoI could be used to further constrain the planetary  interior \cite[e.g.,][]{HELLED2011440}.

\begin{figure}
    \includegraphics[width = 0.495\textwidth]{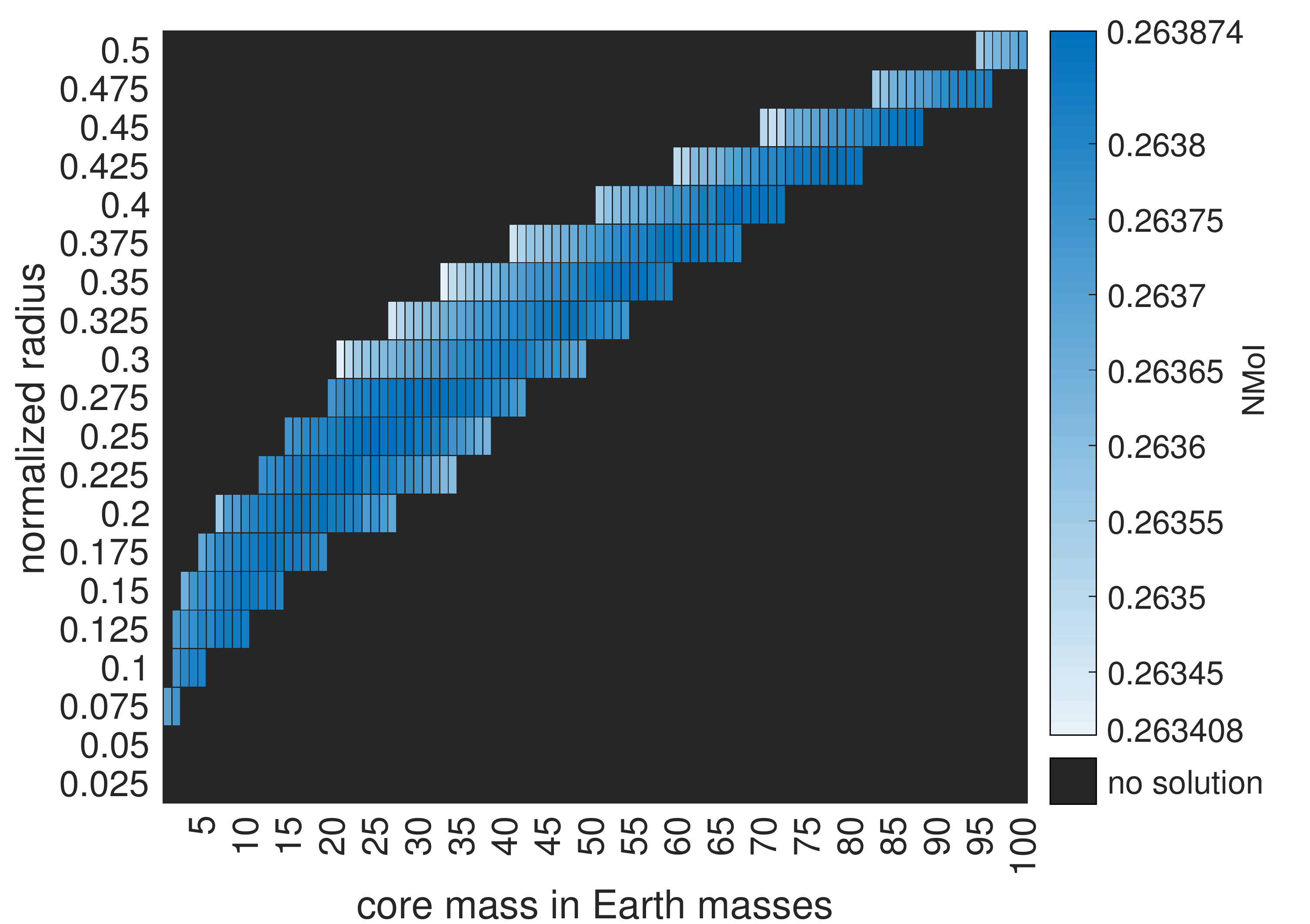}
    \caption{The investigated core properties. Each core property combination of $m\sub{core}$ \& $r\sub{core}$ is either colored according to the inferred MoI value or black, if no solution is found. 
    }
    \label{fig:jupiter_area_plotting_MoI}
\end{figure} 

Figure \ref{fig:jupiter_area_plotting_MoI} shows the relationship between the MoI, core size, and core mass, in our piecewise-polytrope models.
For many core configurations either no \textit{good result} is found or the solution is excluded because it exceeds $P\sub{max}$ or $\rho\sub{max}$ (see section \ref{sec:methods}). Note that especially for small and heavy and for large and light cores no \textit{good results} are found. This is fairly intuitive; the former combination gets restricted by $\rho\sub{max}$ and the latter might produce negative density jumps at the core-envelope boundary ($\rho\sub{core}<\rho\sub{envelope}$). As a result, a large area of the core property space can be excluded by basic physics, before being constrained further by $J_2$ and $J_4$. However, the boundaries of the ``no solution''-area have to be treated with caution; it is possible that some solutions are missed by the optimization algorithm getting stuck in a local minimum. 
Of the core configurations that support valid solutions, light and small cores (left lower) are consistent with the traditional notion of a compact, pure heavy-element core. Solutions in the upper right are consistent with the idea of a diluted core \cite[e.g.,][]{2017Wahl}.

Although our solutions fit the measured $J_2$ and $J_4$-values within their relative uncertainty of $10^{-5}$ and $10^{-4}$, respectively (see table \ref{tab:jupiter_properties}), the relative variation in the MoI is of the order of $10^{-3}$. This suggests that the one-to-one correspondence between $J_2$ and the MoI (eq. \ref{eq:RadauDarwin}) can be broken with sufficiently precise measurements. 
The additional information stored in the MoI, with respect to $J_2$ and $J_4$, can be used to further constrain the core properties (see section~\ref{subsection:MoI_vs_core}) and/or the pressure regime of the density discontinuity in the envelope (see section~\ref{subsection:Transition_behavior}). 

\subsection{The relation between the MoI and the innermost (core) region} \label{subsection:MoI_vs_core}

We suggest that the MoI can be used to further constrain Jupiter's core properties. 
For example, a measurement indicating a large MoI value ($\text{MoI}\gtrsim{0.26355}$) would allow a large variety of core properties. But a smaller one rules out solutions with compact and distinct cores smaller and less massive than $r\sub{core} \lsim{0.3}$ and $m\sub{core} \lsim{20}\unit{M}_\oplus$, respectively. See appendix \ref{sec:MoI_core_relation} for a more detailed treatment of the relation between the MoI and $m\sub{core}$, $r\sub{core}$ and $P\sub{1,2}$.
To be diagnostic, an independently measured MoI value must come with a relative uncertainty not larger than $0.1\%$.
There are different methods to measure and estimate the MoI, e.g., measuring Jupiter's pole precession or the Lense-Thirring acceleration of the Juno spacecraft \cite[e.g.,][]{HELLED2011440}.

\subsection{The relation between the MoI and the density discontinuity in the envelope} \label{subsection:Transition_behavior}
As discussed previously, most density discontinuities occur between $r_{1,2} \approx 0.75-0.9$ (see figure \ref{fig:all_in_one_MoI}). 
Diluted cores tend to have the discontinuity deeper ($r_{1,2} \lsim{0.8}$) and also have smaller MoI values.
Solutions with density discontinuities higher in the envelope ($r_{1,2} \gsim{0.8}$) tend to have large MoI values. 

\begin{figure}
    \includegraphics[width = 0.5\textwidth]{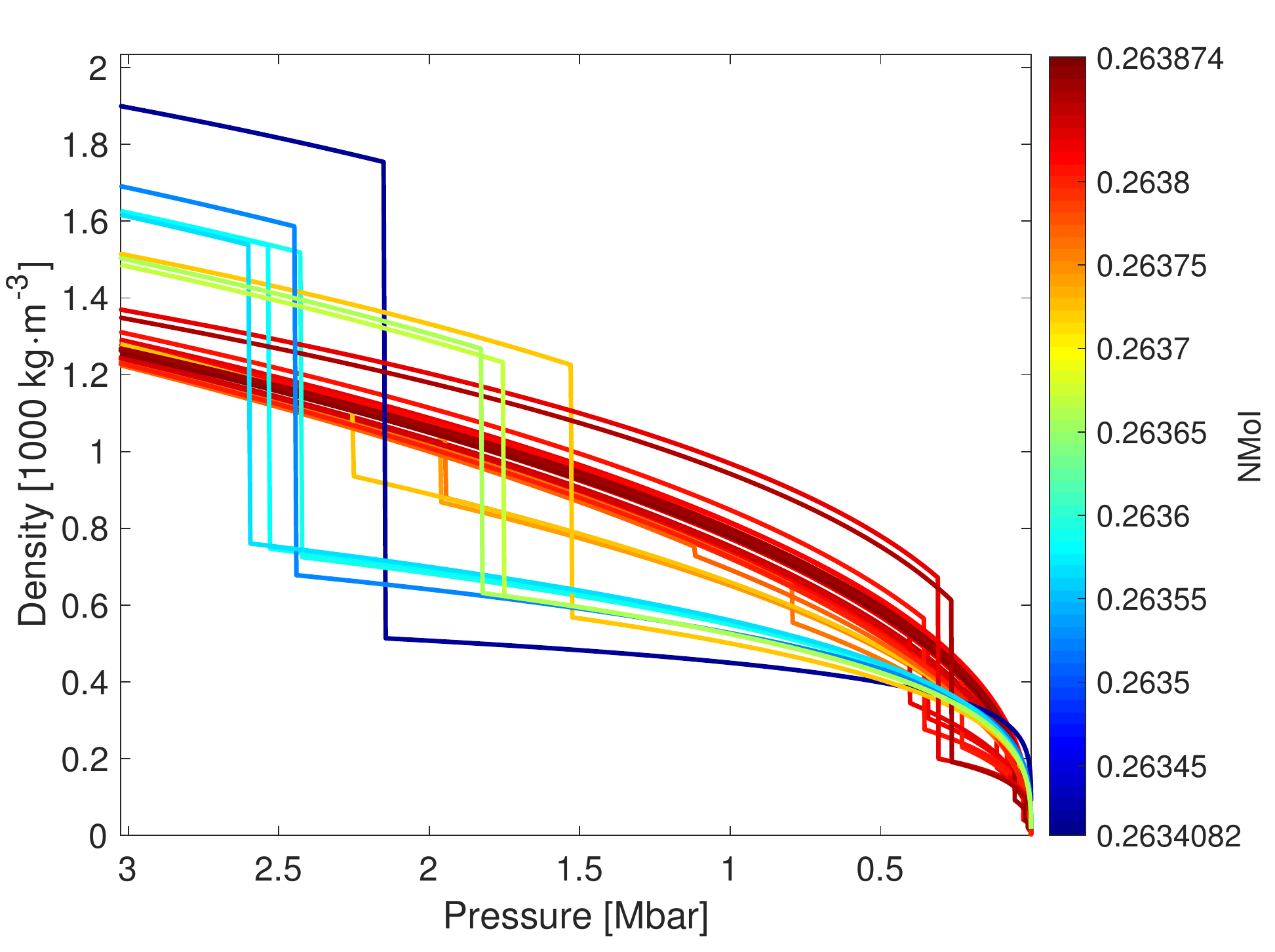}
     \caption{Jupiter's density vs.~pressure in the region of $r_{1,2}$ of a representative selection of \textit{good results}.
    The color indicates the MoI value.
    Diluted cores with low core densities have larger values of $1.5 \lsim{p_{1,2}} \lsim{}3\unit{Mbar}$; lower transition pressures are coupled to higher MoI values and more compact cores.
    } 
    \label{fig:all_in_one_rhofP}
\end{figure}

Figure \ref{fig:all_in_one_rhofP} shows the transition density depending on the transition pressure. The color represents the inferred MoI value. Many density discontinuities occur at transition pressures of $P\sub{trans}=P\sub{1,2} \sim 0.5 - 3 \text{~Mbar}$. This pressure range includes the expected pressure where  hydrogen metallizes at Jupiter's conditions \citep[e.g.,][]{Mazzola2018} and the pressure at which helium is expected to separate from hydrogen \citep[e.g.,][]{Morales2013,Schottler2018}. 
Also there is a clear color trend: diluted cores have lower MoI values ($\lsim 0.2636$) and transition pressures around $1.5 \lsim{P_{1,2}} \lsim{} 3\unit{Mbar}$. Lower values of the transition pressure $P_{1,2}\lsim1.3\unit{Mbar}$ are coupled to higher MoI values ($\gsim{0.2638}$), allowing for more compact cores (see appendix \ref{sec:MoI_core_relation} for further details).

Since a density discontinuity in Jupiter's envelope is typically associated with helium separation from hydrogen, identifying the location of this transition can be linked to the behaviour of hydrogen and hydrogen-helium mixtures in planetary conditions \citep[e.g.,][]{Helled2020}. Therefore, an accurate measurement of Jupiter's MoI could also be linked to the H-He phase diagram.   

\begin{figure}
    \centering
    \includegraphics[width = 0.5\textwidth]{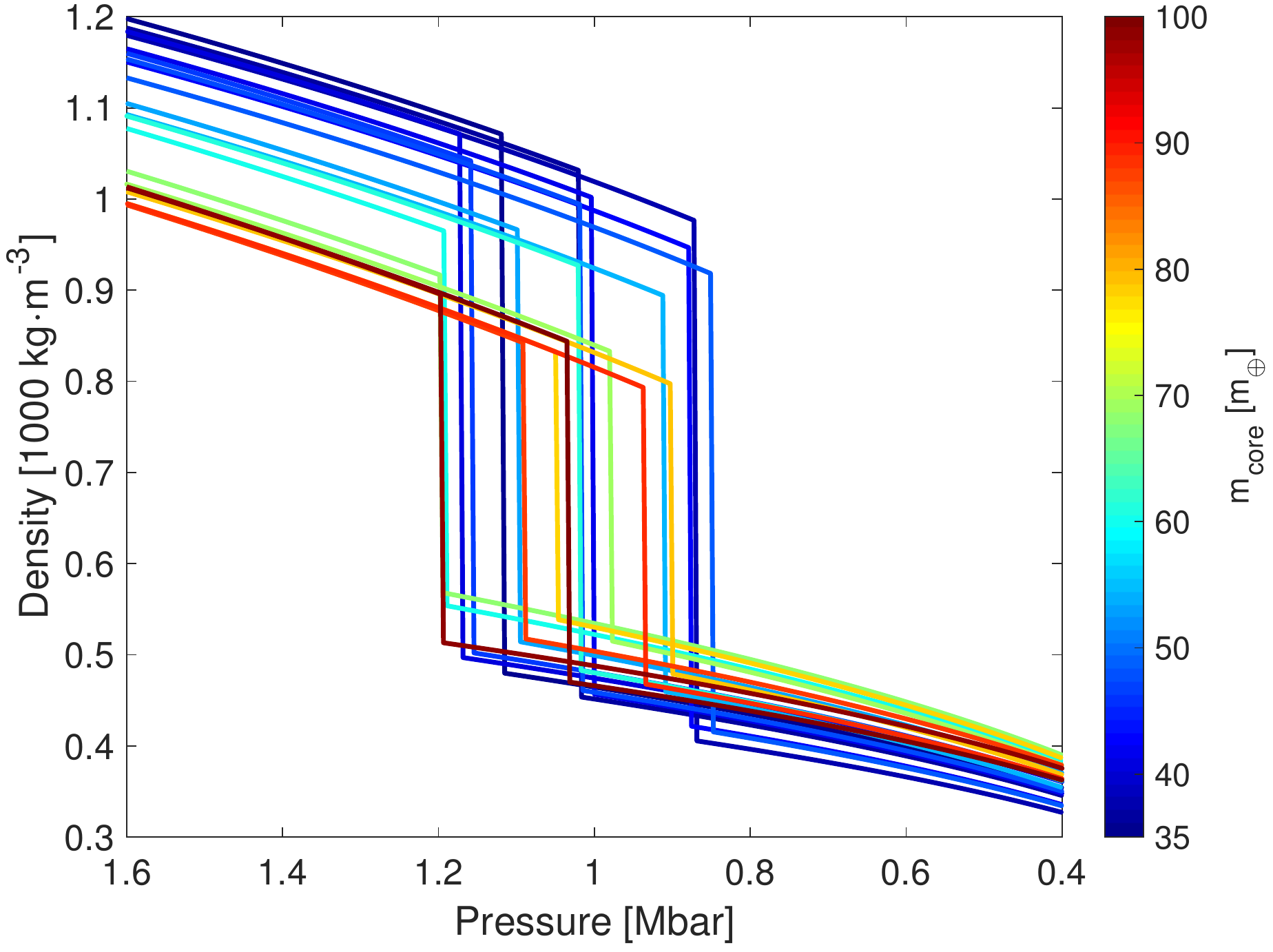}
    
    \includegraphics[width = 0.5\textwidth]{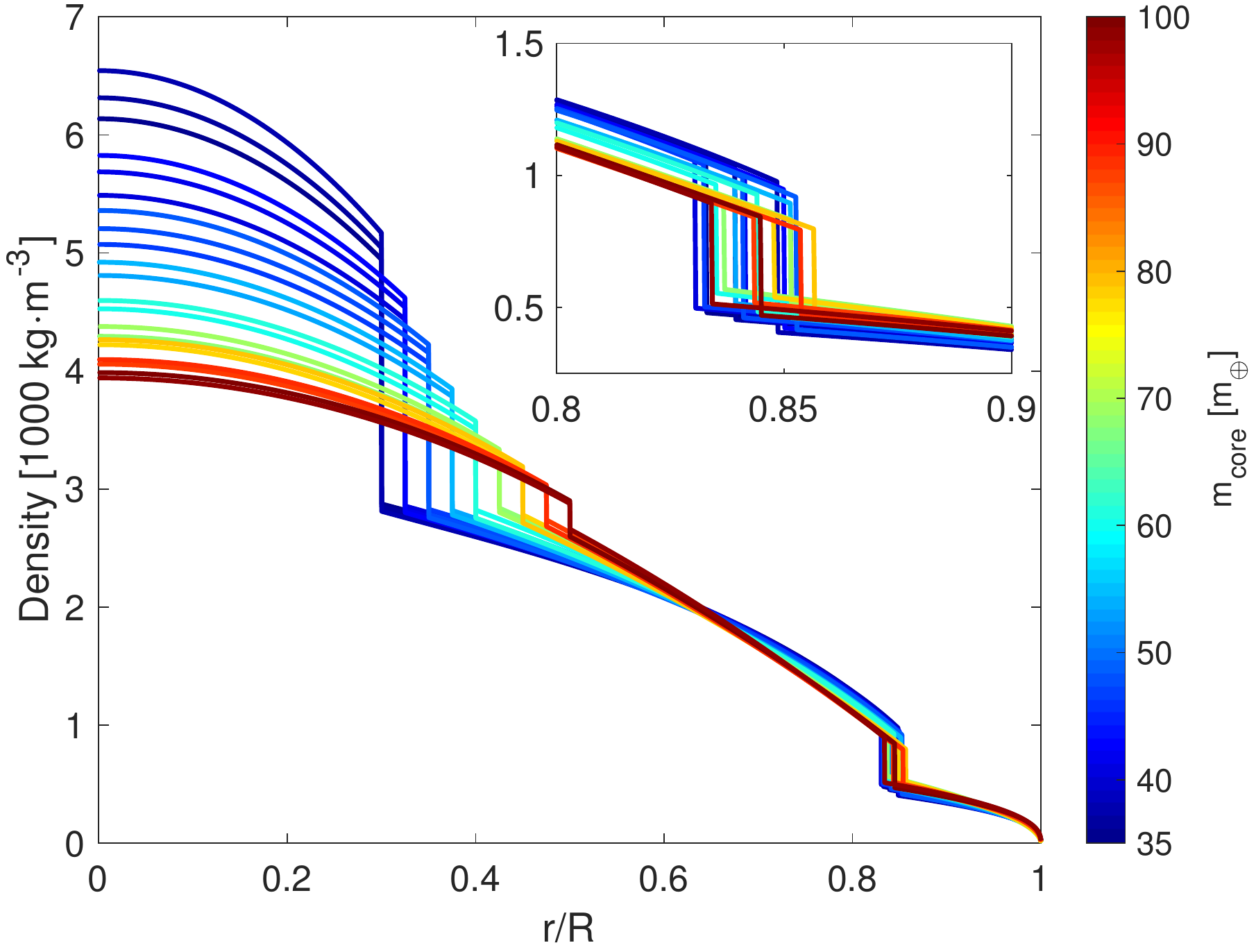}
    \caption{Density profiles of solutions with transition pressures of $P_{\sub{trans}} = 0.8-1.2~\text{Mbar}$. The upper panel shows the density-pressure relation. The lower panel shows the density profile depending on the normalized radius. In both cases the color indicates the core mass of the solution.}
    \label{fig:P_trans_1_Mbar}
\end{figure}

It is interesting to note that our models with 
$P_{\sub{trans}} \sim{1}~\text{Mbar}$, as expected from the hydrogen-helium phase diagram \citep[e.g.,][]{Morales2013,Schottler2018}, have diluted cores. 
If one interprets the models with a density discontinuity around 1 Mbar as being ``more physical'', then this could be a support for a fuzzy core in Jupiter. 
Figure \ref{fig:P_trans_1_Mbar} shows the subset of models with a discontinuity in the envelope between 0.8 Mbar and 1.2 Mbar (corresponding to a transition radius of $r\sub{trans}\sim{0.83-0.86}$). The upper (lower) panel shows the density against the pressure (normalized radius). The color indicates the ``core'' mass of the solution. The density profiles indicate relatively low internal densities of $\rho\sub{core} = 4 - 6.5~\text{kg}\cdot \text{m}^{-3}$ and corresponding core pressures of $P\sub{core} = 36 - 48~\text{Mbar}$, respectively.
The core sizes are found to be between $r\sub{core} = 0.3 - 0.5$ with core masses between $m\sub{core} = 35 - 100~\text{M}_\oplus$, which is rather consistent with an extended dilute core for Jupiter. 
The large magnitude of the density discontinuity could indicate a barrier to convection in this region \citep{Stevenson1977b}, leading to a metal enrichment (depletion) in the inner (outer) layer that contributes to the change in density. 

\subsection{The MoI of a discretized density profile}
This work focuses on trends in the MoI value, and how they relate to other features of the interior. The numerical values themselves, shown in figure 3, shows that the inferred MoI range of ($0.2634 < \text{MoI} < 0.2639$) does not overlap with the suggested MoI values of \cite{2017Wahl} ($0.2640 < \text{MoI} < 0.2644$). This might be surprising given that our empirical models are supposed to cover a large range of possible interior profiles, including approximations of those published models. We believe that, in fact, they do. The apparent discrepancy in MoI value is due not to a material difference between the interior models (i.e., the actual density profiles) but to small differences in calculations involving the discretized versions of the density. Since structure models have finite resolution, one may expect any quantity that involves an integral of density in the radial direction to propagate a discretization error $\epsilon_r=O(1/N)$, where $N$ is the number of specified density values along the radius of the planet. This is especially true in the presence of sharp discontinuities in $\rho(r)$. We verified by comparison with other, EoS-based models (T.~Guillot, private communication) that small differences in the way that these discontinuities are handled, as well as models with lower resolution than used in this study, indeed change the MoI value enough to explain the apparent discrepancy.  

We therefore suggest that the higher MoI values reported previously might be affected by the numerical calculations.  
This should be explored further and resolved, either by agreeing on a consistent method of representing discretized density profiles or by using high enough resolution such that $\epsilon_r$ becomes unimportant. 
Such an analysis is particularly important if accurate measurements of Jupiter's MoI become available and we plan to address this topic in a followup study.   
But regardless of what digit the average MoI ends up showing in the forth decimal place, the \emph{trends} shown in figure \ref{fig:jupiter_area_plotting_MoI} would persist.

\begin{figure*}
    \centering
    \includegraphics[width = 0.5\textwidth]{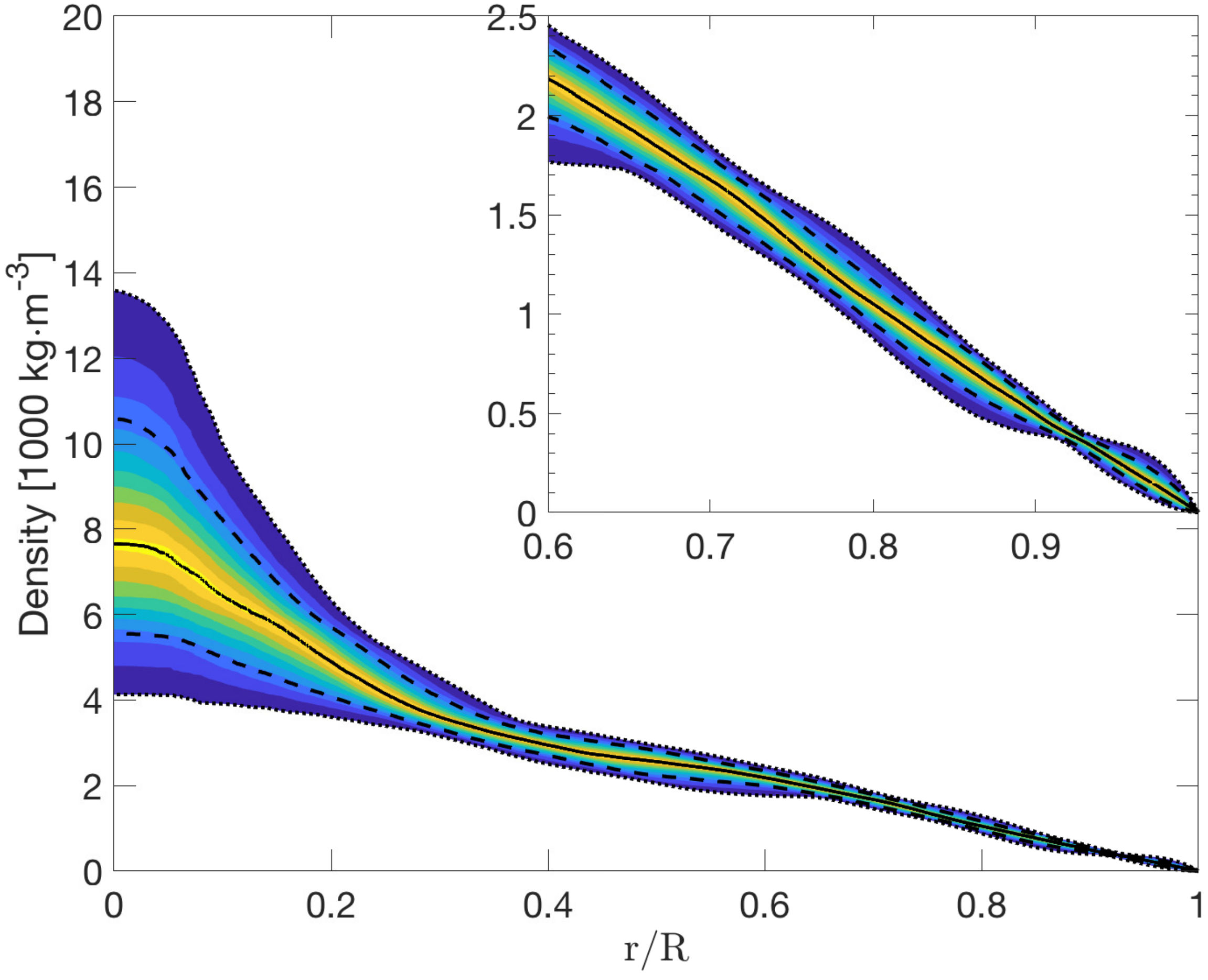}%
    \includegraphics[width = 0.5\textwidth]{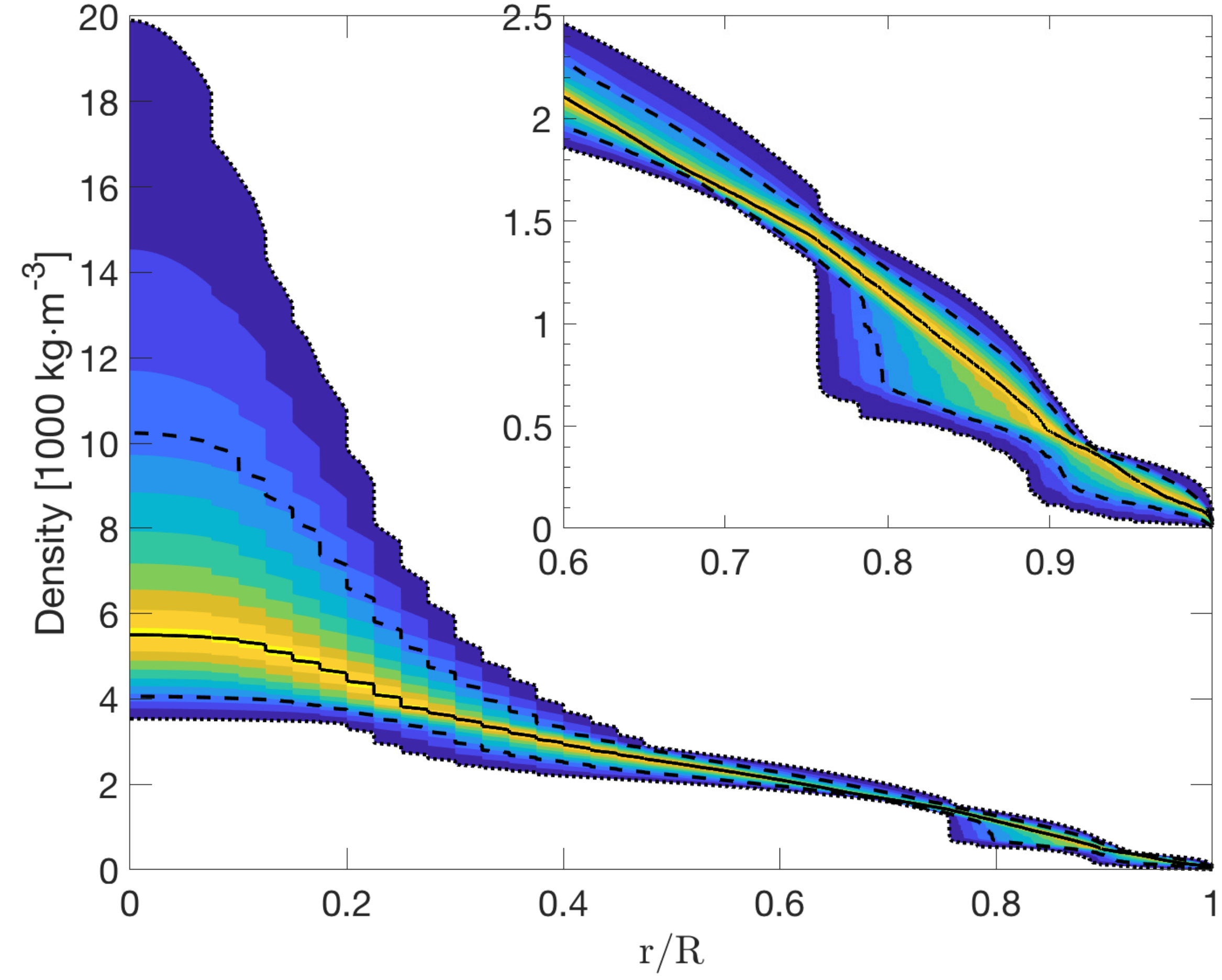}
    \caption{Distribution of density profiles for Jupiter based on $8^{\text{th}}$-degree polynomials (left panel) and polyropes (right panel). The black line marks the sample-median and the dashed lines the width of the one-sigma deviation. The color visualizes the sample distribution and comprises $\sim96\%$ of all solutions. Jupiter's mass, equatorial radius, rotation period, small parameter $m$ and used $J$-values are listed in table \ref{tab:jupiter_properties}. The polynomial-based profiles allow for up to two density jumps and have the same precision as the polytropic-based density structures.
    }
    \label{fig:polynomials}
\end{figure*}

\subsection{Polytropes vs.~Polynomials} \label{subsection:comparison_to_polynomials}
The empirical density profiles we presented above are based on polytropes. However, it is clear that there are many alternatives. One of which are polynomial-based density structures, that are broadly used in literature as e.g. in \citet{2011Helled, HELLED2011440, Anderson2007,Movshovitz2019}.

\begin{figure}[h!]
    %\centering
    \includegraphics[width = 0.5\textwidth]{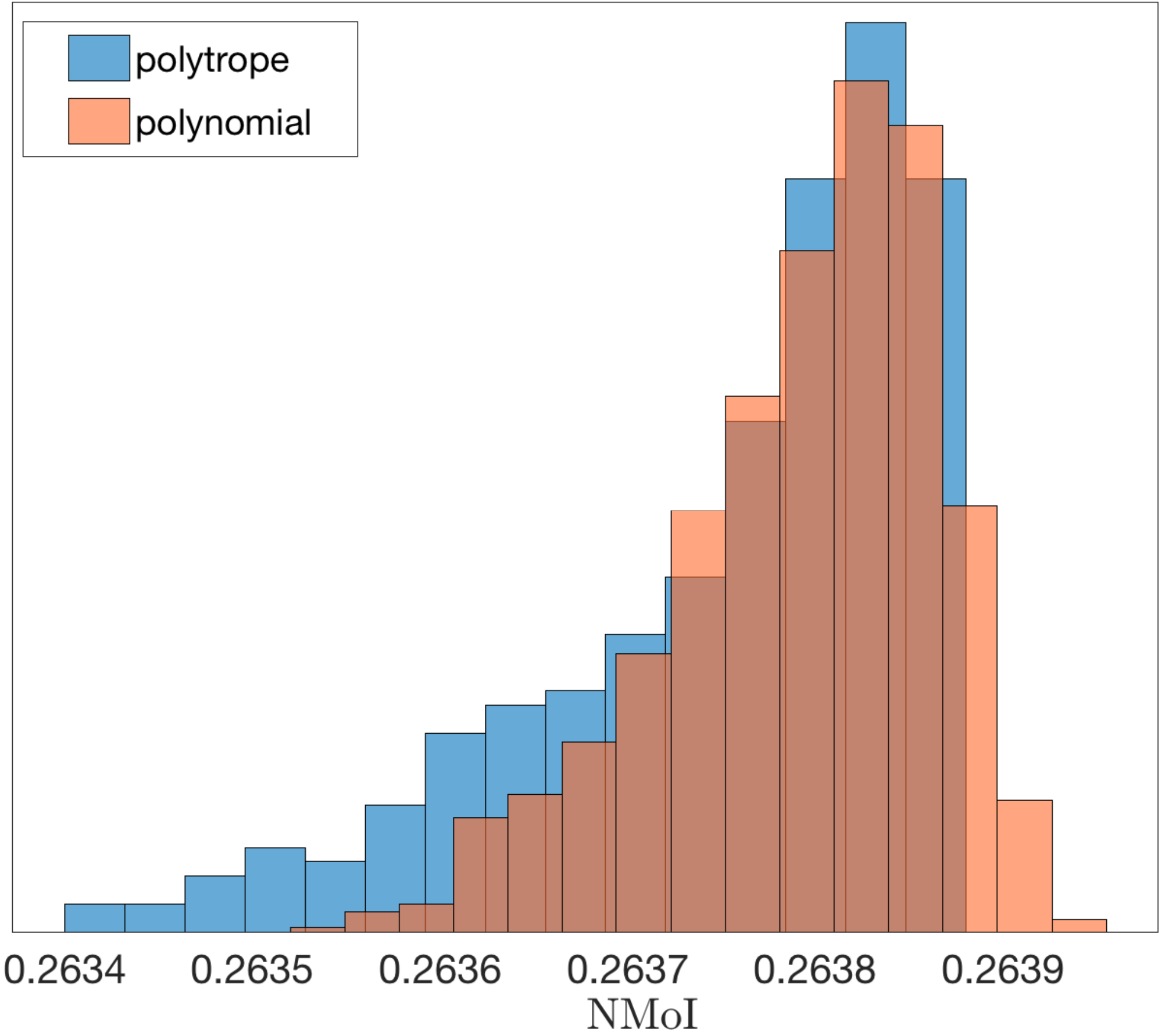}
    \caption{MoI range and distribution of polytrope-based density structures (blue colored) and polynomial-based density structures (red colored). For both modeling methods the ranges and distribution of the MoI almost perfectly overlap.}
    \label{fig:histogram}
\end{figure}
To test and detect potential biases of polytrope-based density profiles, we compare our results (calculated MoI range and density profiles) to a broad range of polynomial-based solutions. For the latter we build on the work of \cite{Movshovitz2019} and represent the density profile using a degree-8 polynomial with up to two density discontinuities superimposed, with the range of interior solutions guided by an MCMC. Further details on the precise parameterization are  described in \metalt. To ensure that differences in the results emerge solely from the different methods, the same planet properties (table \ref{tab:jupiter_properties}) and gravity field calculation method are used for the polynomial-based calculation.

Figure \ref{fig:polynomials} shows the resulting distribution of density profiles vs.~Jupiter's normalized mean radius for both polynomial-based profiles (left panel) and polytrope-based profiles (right panel). The solid black line is the ensemble-median and the dashed line marks the 1-sigma width of all density profiles. The color visualizes the width of the distribution.
The solution space of polynomials-based structure profiles is almost a complete subset of the solution space of polynomial-based density profile.

Figure \ref{fig:histogram} shows the MoI range and its distribution of the polytropes-based (blue colored) and a polynomial-based (red colored) density profiles. The range and distribution of MoI values are almost identical.
The similarities in the MoI range and distribution, and in the density profile solution space, increase our confidence in the choice of polytropes to represent the interior pressure-density structure, and in the above inferences.

Both representation of the density profiles have limitations. 
For example, polytropes typically overestimate Jupiter's density in the atmosphere as measured by the Galileo Entry Probe. Admittedly, this effect is supposed to be small as it only concerns Jupiter's outermost 0.5 M$_\oplus$ and its surface density is barely captured by $J_2$ and $J_4$ but mostly effects higher-order $J$-values. Further, polynomial-based structure models that fully account for Jupiter's measured surface density yield similar results with respect to the MoI range and density profiles. Finally, the Galileo entry probe could only resolve one small spot at Jupiter's dynamical atmosphere, hence an extrapolation to its entire atmosphere has to be treated with caution. In addition, the atmospheric density structure is thought to change over time, this puts additional uncertainties to Jupiter's surface structure. 

Polynomials on the other hand are very general but can produce density profiles that seem nonphysical. Therefore, in both cases, a comparison to physical models would be useful and can be used to exclude some of the solutions. 

\section{Summary and Conclusions}\label{sec:summary}
We present new empirical density profiles of Jupiter. Each density profile is represented by up to three polytropes, and is set to fit Jupiter's mass, equatorial radius, rotation rate, and the recently measured $J_2$ and $J_4$. Clearly, more accurate model evaluations including higher order gravitational harmonics are highly valuable and subject of current research. Nevertheless, since higher-order harmonics are more affected by the dynamics, the results presented here, are expected to remain unchanged. 

First, we infer the connection between the properties of the innermost region of Jupiter, the density discontinuity in the envelope and the inferred MoI value. 
We then investigate the sensitivity of $J_2$, $J_4$ and the MoI to various core properties. Next we explore under what condition the MoI further constrains Jupiter's internal structure.  We also compared our polytrope-based structure models to polynomial-based models. 

While it is possible that Jupiter's density profile could be tightly, or perhaps fully, constrained by using many gravitational moments to high precision (leaving no additional information to be found in the MoI), in practice an accurate independent determination of the MoI is more feasible. Especially since the measured values of high-order gravity coefficients are increasingly ``contaminated'' by dynamic effects. It is possible that the Juno extended mission will be able to provide this measurement. Even if this measurement comes with relatively large uncertainty, it would still be very valuable to compare the measured value to the one inferred by structure models.

Our main conclusions can be summarized as follows:
\noindent
\begin{itemize}[
  align=left,
  leftmargin=1.4em,
  itemindent=0pt,
  labelsep=0pt,
  labelwidth=1.4em
]
\item 
We confirm that the MoI contains additional information in comparison to the gravitational coefficients $J_2$ and $J_4$. 

\item Jupiter's MoI value ranges from $0.263408 - 0.263874$ giving relative (absolute) changes in the order of $10^{-3}$ ($10^{-4}$).
Therefore, we suggest that if Jupiter's MoI is accurately measured (with an uncertainty smaller than 0.1\%) 
it can further constrain Jupiter's internal structure. 
\item Models with a transition pressure of $\sim1$~Mbar, as expected from the H-He phase diagram, indicate a fuzzy  core for Jupiter with sizes between 30-50\% of the planet's radius, consisting up to 30\% of its total mass. 
\item Our results are independent on the used density profile representation of polytropes and are the same when using $8^{\text{th}}$-order polynomials. 
\end{itemize}
We suggest that empirical structure models can be used to further understand Jupiter's interior. In the future, the inferred density profiles, which provide the density-pressure relation in Jupiter should be interpreted in terms of composition and its depth dependence using physical equations of state and we hope to address this topic in future research. 

\acknowledgments
We thank the refree for valuable comments that helped to improvde the manuscript.
We also thank D.~Stevenson, T.~Guillot and the Juno science team members for valuable discussions. 
R.H.~acknowledges support from SNSF grant 200021\_169054. 
J.J.F.~acknowledges the support of NASA grant NNX16AI43G, NSF AST grant 1908615, and University of California grant A17-0633-001 to the Center for Frontiers in High Energy Density Science. We acknowledge use of the lux supercomputer at UC Santa Cruz, funded by NSF MRI grant AST 1828315.

\appendix

\section{Code Validation}

Here we show that our calculation method can reproduce well-known solutions. 
First, we evaluate the MoI of a non-rotating planet represented by only one polytrope and compare it to the published results of \cite{Lattimer_2001}. Table \ref{tab:comp.MoI} lists the MoI values for various index values ($n$-values in eq. \ref{eq:polytrope}), evaluated by ToF, and compares it to the solution by \cite{Lattimer_2001}. The third column shows the relative difference between the solutions. The relative error agrees with the method's precision.

\begin{figure}
    \centering
    \includegraphics[width = 0.5\textwidth]{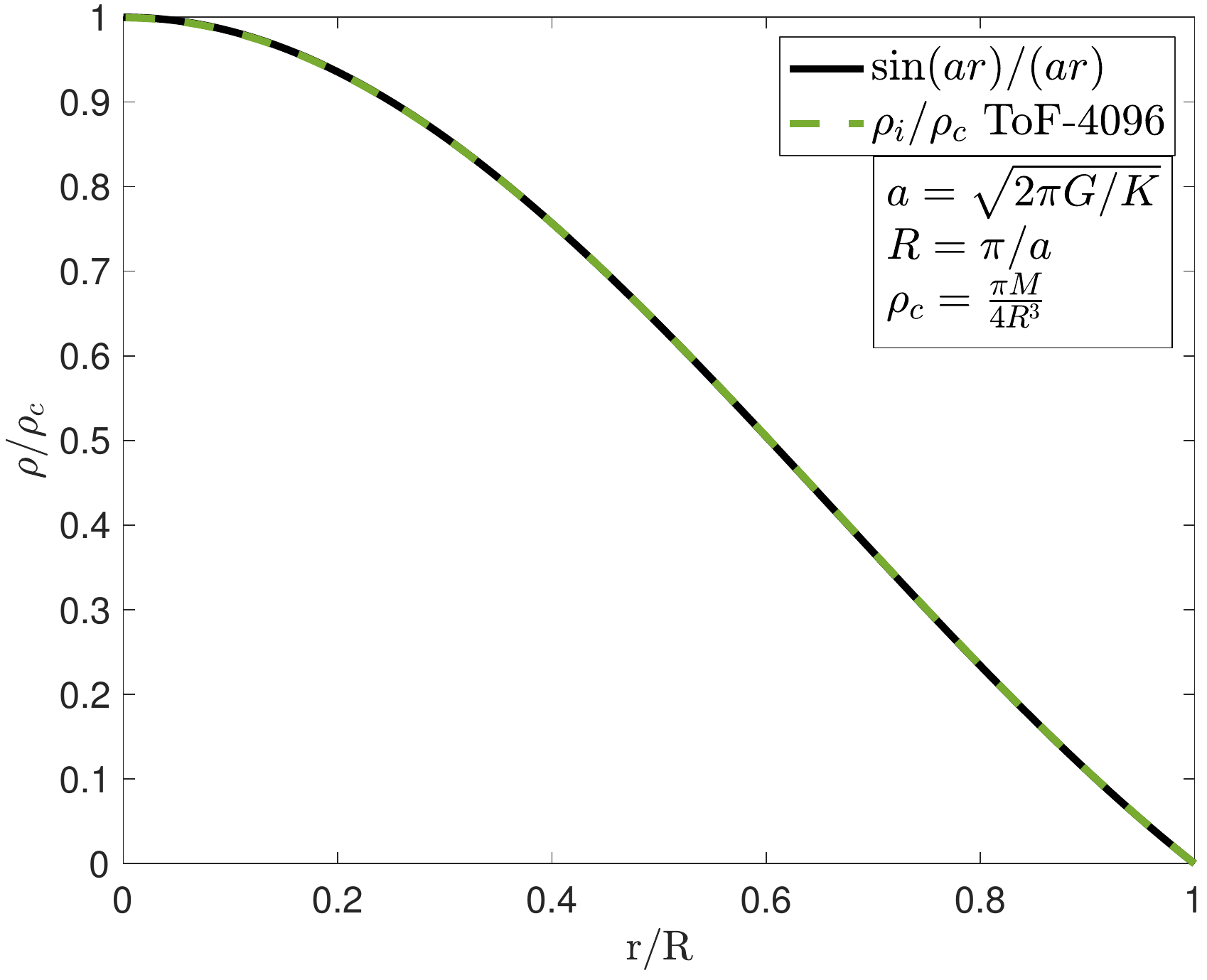}
    \caption{Normalized density profiles of a non-rotating index-1-polytrope. The dashed green line marks the density distribution evaluated by ToF. The solid black line marks the density profile of the corresponding analytical solution. The lines are almost perfectly overlapping and only deviate within the order of $10^{-4}$.}
    \label{fig:test_non_rot_density_profile}
\end{figure}

Second, the density profile of a non-rotating index-1-polytrope is evaluated by ToF and compared to its analytical solution. Figure \ref{fig:test_non_rot_density_profile} shows the normalized density vs.~the normalized planetary radius. The black line marks the density distribution of the analytic solution while the dashed green line represents the density profile as evaluated by ToF. The shape function of 4096 equipotential levels are explicitly calculated. Normalization factors are given in the figure, where $G$ is the gravitational constant, $K$ the polytropic constant and $M$ the total mass of the planet. Note that both solutions are almost identical.

Finally, $J_2$ and $J_4$ of an index-1-polytrope, evaluated by different methods, are compared. Table \ref{tab:comp.J2J4} lists $J_2$ and $J_4$ values as evaluated by ToF (4096 equipotential layers), CMS (512 layers) by \cite{Hubbard_2013}, the exact Bessel solution (eBe) and Consistent Level Curve (CLC), both by \cite{Wisdom_2016}. Relative differences in $J_2$ and $J_4$ between ToF and the other methods are within the method's relative precision of $10^{-4}$.

\begin{table}
\centering
 \begin{tabular}{c| c c c}
 \hline
$n$ & ToF & \cite{Lattimer_2001} & rel. difference \\
 \hline
0.5 & 0.32587 & 0.32593 & 1.72$*10^{-4}$ \\
1.0 & 0.26139 & 0.26138 & 3.83$*10^{-5}$ \\
2.0 & 0.15497 & 0.15485 & 7.58$*10^{-4}$ \\
\hline
 \end{tabular}
\caption{The MoI of a non-rotating polytrope, evaluated by ToF (left column) or proposed by \citet{Lattimer_2001} (middle column) for various polytropic indices. Relative differences in the MoI values are shown in the third column. It is found that the relative precision of our ToF method does not exceed $10^{-4}$.}
\label{tab:comp.MoI}
\end{table}

\begin{table}
\centering
 \begin{tabular}{r| c c c c}
 \hline
 & ToF & CMS-512 & eBe & CLC \\
 \hline
$J_2*10^2$ & 1.39955 & 1.39892 & 1.39885 & 1.39885 \\
$-J_4*10^4$ & 5.32240 & 5.31880 & 5.31828 & 5.31828 \\
\hline
 \end{tabular}
\caption{$J_2$ and $J_4$ of an index-1-polytrope evaluated by either ToF, CMS (by \cite{Hubbard_2013}), the exact Bessel solution (eBe) and Consistent Level Curve (CLC) both by \cite{Wisdom_2016}. Relative differences between our ToF method and the other methods are not larger than $10^{-4}$.}
\label{tab:comp.J2J4}
\end{table}

\section{All inferred density profiles} \label{sec:all_inferred_profiles}
\begin{figure}
    \includegraphics[width = 0.5\textwidth]{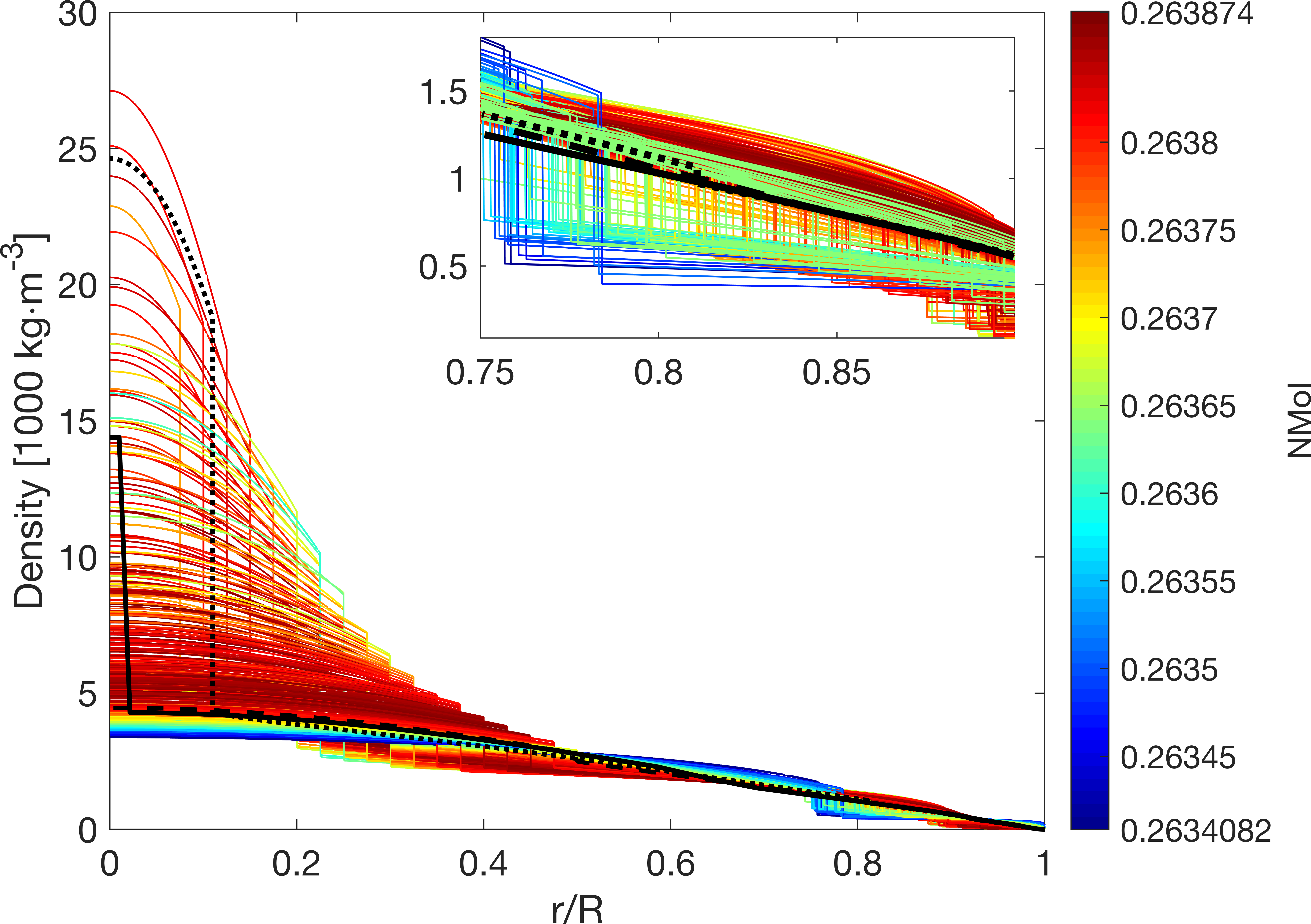}
        \caption{Jupiter's density vs.~normalized radius of all \textit{good results}. The color of each solution illustrates its MoI value. For comparison, published results of \cite{Debras_2019} (black solid line), \cite{2017Wahl} (black dashed line) and \cite{Miguel2016} (black dotted line) are included.}
    \label{fig:all_in_one_MoI_all}
\end{figure}

\begin{figure}
    \centering
    \includegraphics[width = 0.5\textwidth]{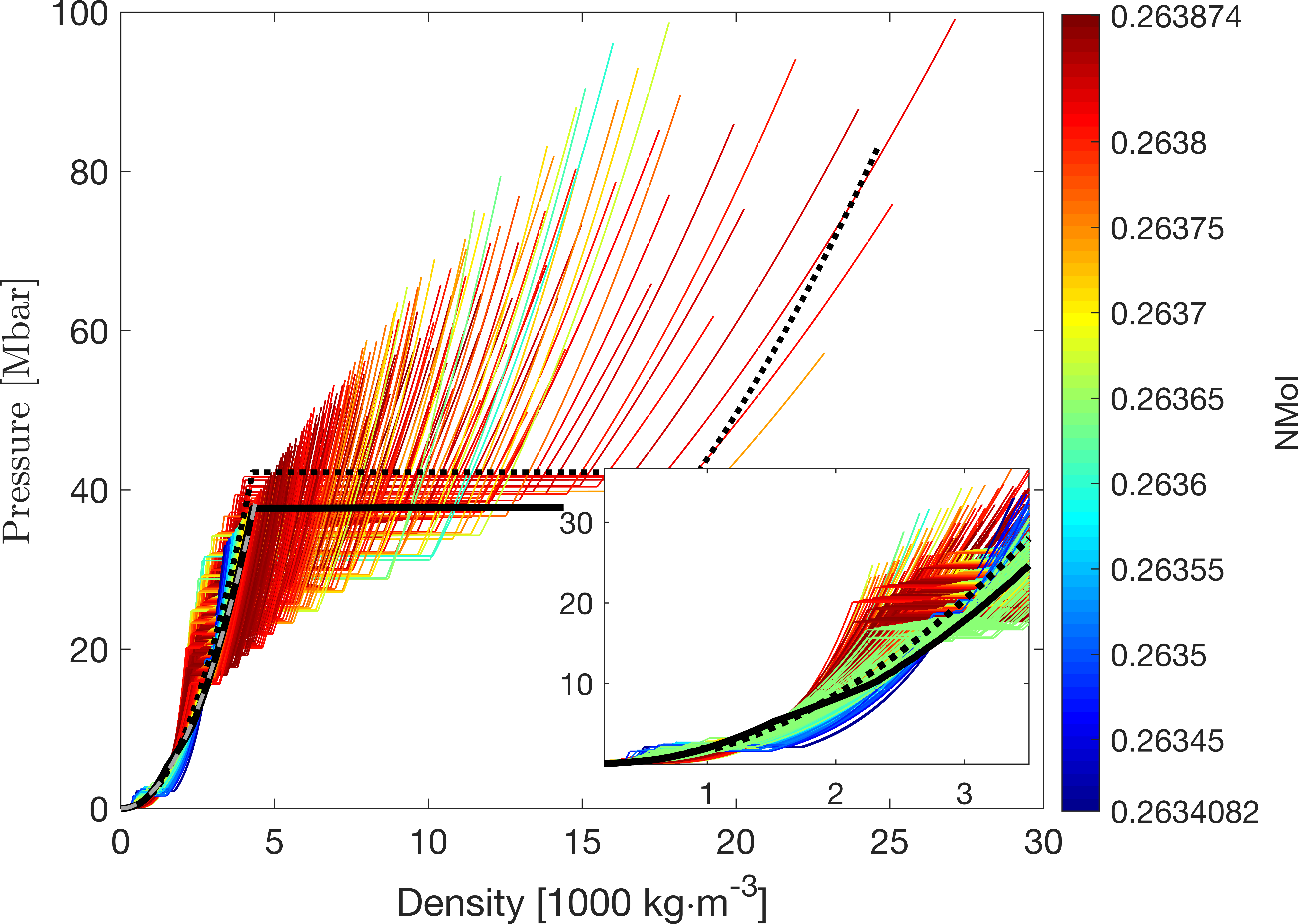}
    \caption{Pressure vs.~density of all \textit{good results}. The colors indicates the MoI of each solution. The solid and dotted black curve shows solutions of \cite{Debras_2019} and \cite{Miguel2016}, respectively. The dashed grey line marks the solution of an index-1-polytrope.
    }
    \label{fig:all_in_one_P_of_rho}
\end{figure}

Figure \ref{fig:all_in_one_MoI_all} shows \textbf{all} \textit{good results} density profiles of Jupiter. The color is representing the MoI-value of each solution. The solid, dotted and dashed black lines represent solutions of \cite{Debras_2019}, \cite{2017Wahl} and \cite{Miguel2016}, respectively.
Figure \ref{fig:all_in_one_P_of_rho} shows the pressure vs.~density of \textbf{all} \textit{good results} of Jupiter. The solid and dotted black lines mark solutions of \cite{Debras_2019} and \cite{Miguel2016}, respectively. The dashed grey line shows the solution of an index-1-polytrope. Obviously the external profiles are in agreement with our solution space, although the result of \cite{Debras_2019} clearly marks an upper (lower) pressure bound at a density of $\sim{1500}$~kg$\cdot$m$^{-3}$ ($\sim{2800}$~kg$\cdot$m$^{-3}$). \\

\section{Constraining power of the MoI} \label{sec:MoI_core_relation}
Table \ref{tab:MoI:ranges} shows the constraining power of the MoI on $r\sub{core}$, $m\sub{core}$ and $P\sub{trans}$ for very distinct ranges of MoI.
The upper two parts of table \ref{tab:MoI:ranges} tabulate the results of figure \ref{fig:jupiter_area_plotting_MoI} and list MoI ranges for different core properties. The lower part describes the relation between the transition pressure $P\sub{1,2}$ and the MoI. It emphasizes the importance of having an independent measurement of Jupiter's MoI. Note that this table is supposed to be a look-up-table for a future measured MoI.
However, it also allows to compare our results to comparative studies. 
\begin{table}
\caption{MoI ranges depending on the size of the core (top part), mass of the core (mid part) and the transition pressure at the density discontinuity in the envelope (lower part). We mark the changing digits of interest in bold.}
\centering
\begin{tabular}{r @{}>{${}}c<{{}$}@{} l | r @{}>{${}}c<{{}$}@{} l}

\multicolumn{3}{c}{MoI range} & \multicolumn{3}{l}{$r\sub{core}$ range} \\
\hline
0.263\textbf{41}&-&0.263\textbf{47} & 0.300&-&0.375 \\
0.263\textbf{47}&-&0.263\textbf{54} & 0.300&-&0.450 \\
0.263\textbf{54} &-& 0.263\textbf{69} & 0.150&-&0.500 \\
0.263\textbf{69} &-& 0.263\textbf{82} & 0.075&-&0.475 \\
0.263\textbf{82} &-& 0.263\textbf{87} & 0.125&-&0.450 \\
\\
\multicolumn{3}{c}{MoI range} & \multicolumn{3}{l}{$m\sub{core}$ range in [M$_\oplus$]} \\
 \hline
0.263\textbf{41} &-& 0.263\textbf{47} & 21&-&41 \\
0.263\textbf{47} &-& 0.263\textbf{55} & 21&-&72 \\
0.263\textbf{55} &-& 0.263\textbf{65} & 7&-&98 \\
0.263\textbf{65} &-& 0.263\textbf{69} & 1&-&100 \\
0.263\textbf{69} &-& 0.263\textbf{77} & 2&-&91 \\
0.263\textbf{77} &-& 0.263\textbf{82} & 3&-&96 \\
0.263\textbf{82} &-& 0.263\textbf{85} & 4&-&86 \\
0.263\textbf{85} &-& 0.263\textbf{87} & 13&-&88 \\
\\
\multicolumn{3}{c}{MoI range} & \multicolumn{3}{l}{$P\sub{trans}$ range in [Mbar]} \\
 \hline
0.263\textbf{41} &-& 0.263\textbf{60} & 1.5 &-& 3 \\
0.263\textbf{60} &-& 0.263\textbf{76} & 0.3 &-& 4.35 \\
0.263\textbf{76} &-& 0.263\textbf{87} & 0.01 &-& 1.3 \\
 \end{tabular}
\label{tab:MoI:ranges}
\end{table}

\section{Constant density core vs. Compressed Core} \label{subsection:CDCvsPC}
Structure models often assume a constant density core (CDC) rather than a compressed core (PC) (e.g. \cite{HELLED2011440}, \cite{Hubbard_2016}, \cite{Ni2018}, \cite{Debras_2019}). This assumption may be inappropriate for compressible materials. 
Here we investigate the change in the $J_{2n}$ and the MoI values when using a CDC vs. PC - represented by a polytrope - in a Jupiter-like planet. This planet is not exactly Jupiter, as its gravity field is different, but has still the same mass, radius and rotation period. To diminish potential effects on $J_{2n}$ and the MoI that are not related to the different core types, we consider only a two-layered density profile (consisting of a core and an envelope) for each core type. For both core models the core mass, core radius and polytropic envelope are the same. Hence the inferred error on $J_{2n}$ and MoI represents the differences between the two core types.

Note that we can only fix either the core mass or the core mean density $\Bar{\rho}\sub{core}$ for both core types, as $M$ and $\Bar{\rho}\sub{core}$ are related via $\Bar{\rho} = M/V$. 
A system with fixed $\Bar{\rho}\sub{core}$ and $m\sub{core}$ and total mass M (fixed as a requirement) is over-constrained: A different density distribution changes the planetary shape and therefore its volume. As a consequence, we only present the results for a fixed $m\sub{core}$. Fixing the core average density leads to similar conclusions.

\begin{figure}
    \includegraphics[width = 0.5\textwidth]{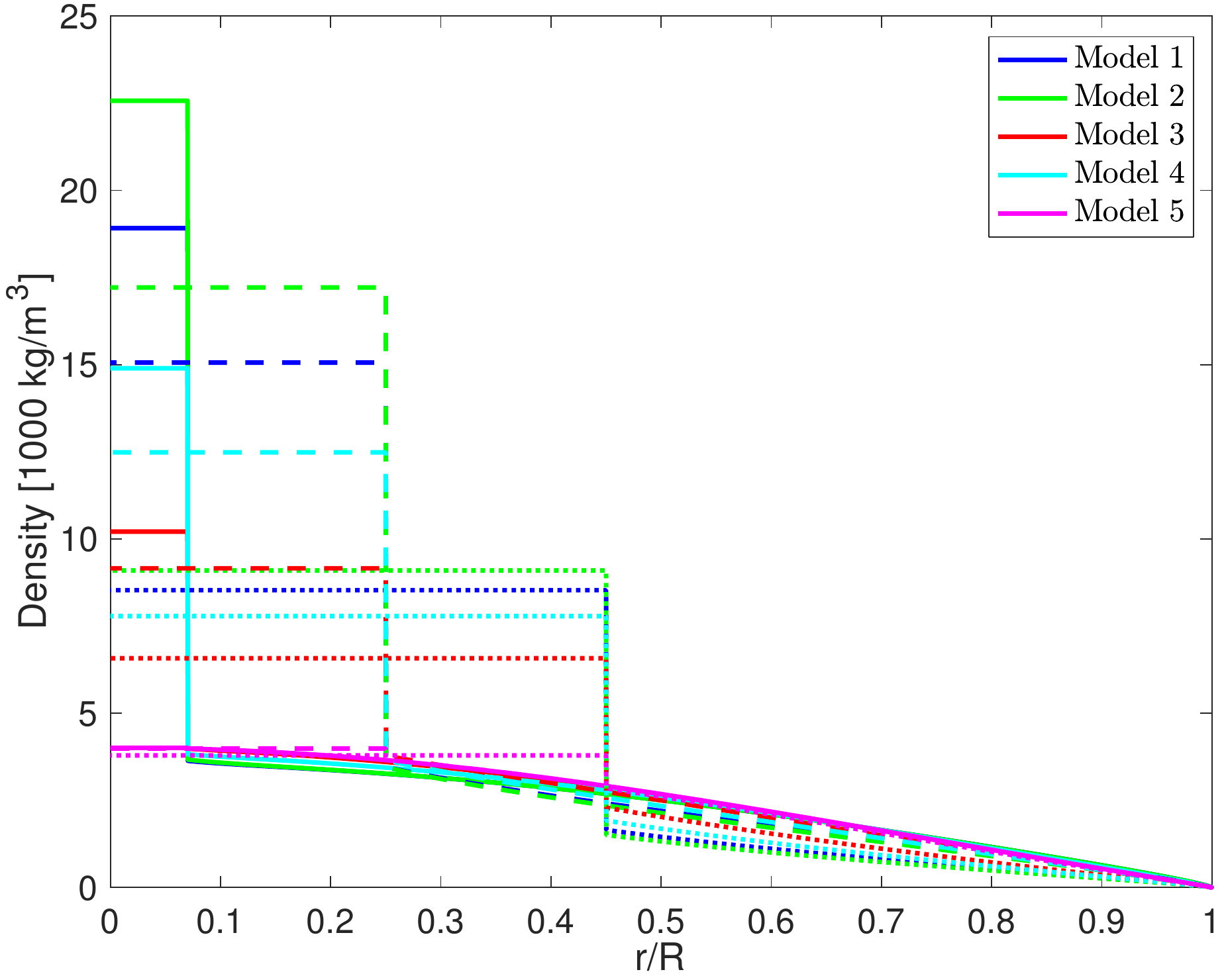}
    \includegraphics[width = 0.5\textwidth]{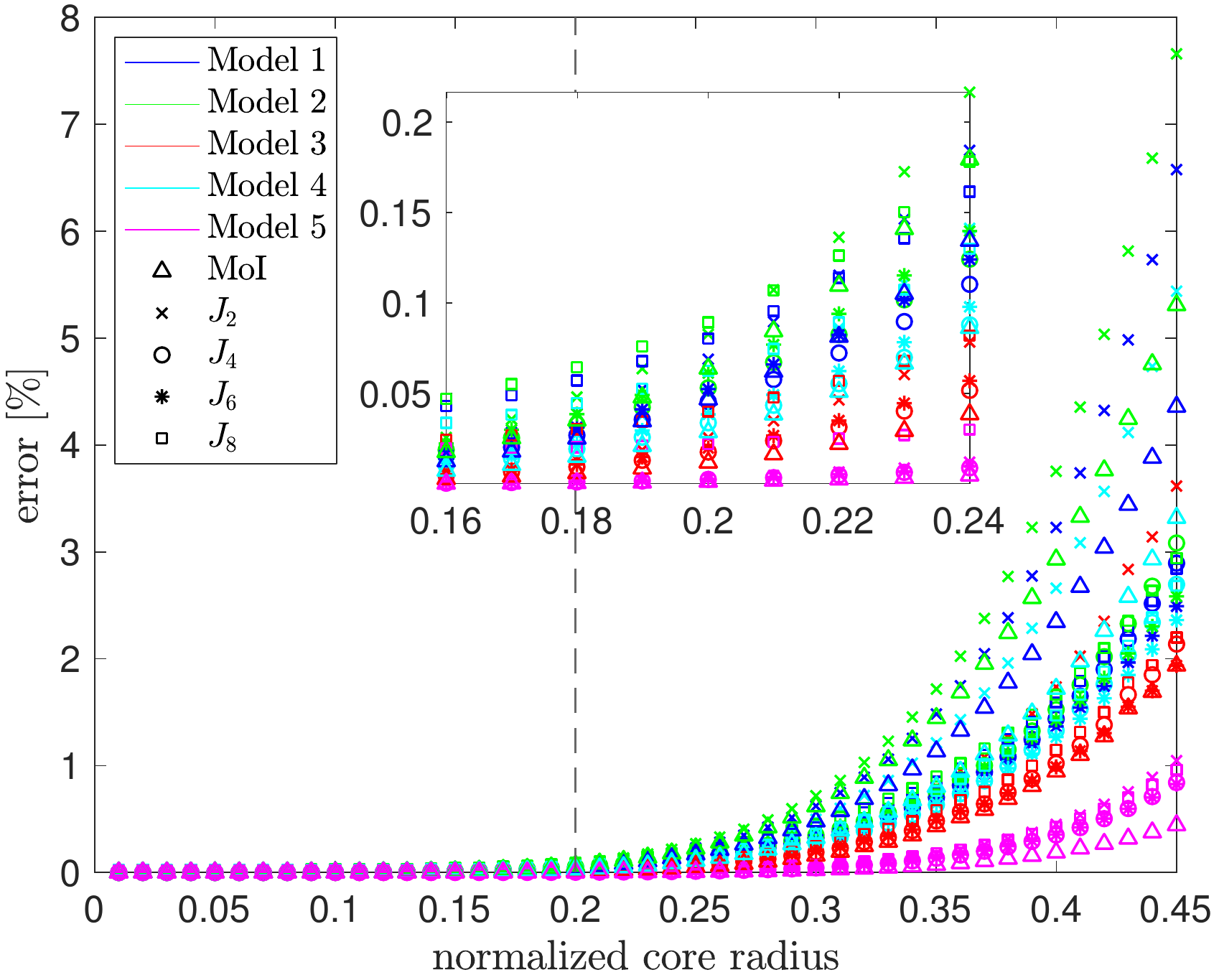}
    \caption{{\bf Top:} The five tested CDC models. To generalize the study, five different core properties and envelope polytropes are tested. The color refers to the different models (only plotted for a CDC) plotted at different core radii: $r\sub{core} = 0.07\text{ } (\text{solid lines}), \text{ }0.25\text{ } (\text{dashed lines}),\text{ } 0.45\text{ } (\text{dotted lines})$. Models 2 \& 5 are most extremes with respect to core mass and density jump at the core-envelope boundary. 
    {\bf Bottom:} Inferred error in the MoI and the $J_{2n}$-values depending on the core size. Behavior of the error (y-axis), representing the differences in various variables arising by replacing a CDC by a PC, depending on the core size (x-axis). The color represents the different models. The different variables ($J_{2n}$ and MoI) are expressed by various symbols. The dashed line at $r\sub{core}=0.2$ marks the maximum core radius whose inferred error is within the method's uncertainty of ToF.}
    \label{fig:Model_comparison}
\end{figure}

To investigate possible effects of the core properties on the inferred $J$-values and the MoI, we consider five different core densities and envelope polytropes. A percentage error is evaluated for different core sizes by using the following equation: 
\begin{equation}
    \text{error} = 100\times \left( \frac{\text{value}_\text{cdc}}{\text{value}_\text{pc}}-1 \right).
    \label{eq:error}
\end{equation}
The top panel of figure \ref{fig:Model_comparison} shows the five models, color-coded and plotted for a CDC at core sizes of $r\sub{core} = 0.07$ (solid lines), $r\sub{core}=0.25$ (dashed lines), and $r\sub{core}=0.45$ (dotted lines). Model 2 represents a massive core, that leads to a large discontinuity at the core-envelope boundary, while Model 5 represents a relatively smooth transitions. The other models represent intermediate cases.

The bottom panel of figure \ref{fig:Model_comparison} shows the percentage differences (denoted as errors) in $J_{2n}$ and the MoI (y-axis) by comparing a CDC with a PC depending on the core size (x-axis). 
The colors represent the models (shown in figure~\ref{fig:Model_comparison}) and the symbols the corresponding parameter. 
Note that for large core radii $J_2$ is affected the most, mainly followed by the MoI. This seems to be consistent with the contribution functions shown in figure \ref{fig:contr_function} and the Radau-Darwin approximation (eq. \ref{eq:RadauDarwin}). However, for small cores, i.e., $r\sub{core} \lsim{0.2}$, the higher-order harmonics are more affected. 
Our interpretation for this behaviour is as follows. 
The gravitational moments are blind to the planet's innermost region (figure \ref{fig:contr_function}). Therefore the inferred errors on the $J_{2n}$ are not generated by the different core types directly. However the core densities of the CDC and the PC are different. This affects the shape and therefore the volume of the whole planet. In return the density profile in the envelope changes. These changes are primarily affecting the higher order gravitational coefficients, due to their relatively high maximal contribution. Hence, for small cores, the inferred errors on the higher order $J_{2n}$-values are larger than the inferred errors on the lower ones.

For core radii larger than a critical core size ($r\sub{crit}$) the direct affect on $J_2$ by the different core types gets dominant. $r\sub{crit}$ depends on the underlying core model (i.e. its exact mass and density). In our models the critical core size is around $r\sub{crit}\sim{0.2}$ and therefore in agreement with the contribution functions of \cite{2011Helled}.

We find that the error is increasing with increasing core size. As a result, the largest acceptable CDC depends on the demanded precision. Since in this paper we use fourth-order ToF that has a relative precision of $\sim 10^{-4}$, the maximal CDC radius should not exceed $r_{\text{cdc}}\lsim{0.2}$.

Overall, we find that differences between a CDC and a PC strongly depend on the actual core properties. For example, Model 2, which has the highest considered core mass, produces the largest error, in contrast to the smooth (diluted) core of Model 5 (blue and purple symbols, respectively, in figure~\ref{fig:Model_comparison}).

\begin{figure}
    \centering
    \includegraphics[width = 0.5\textwidth]{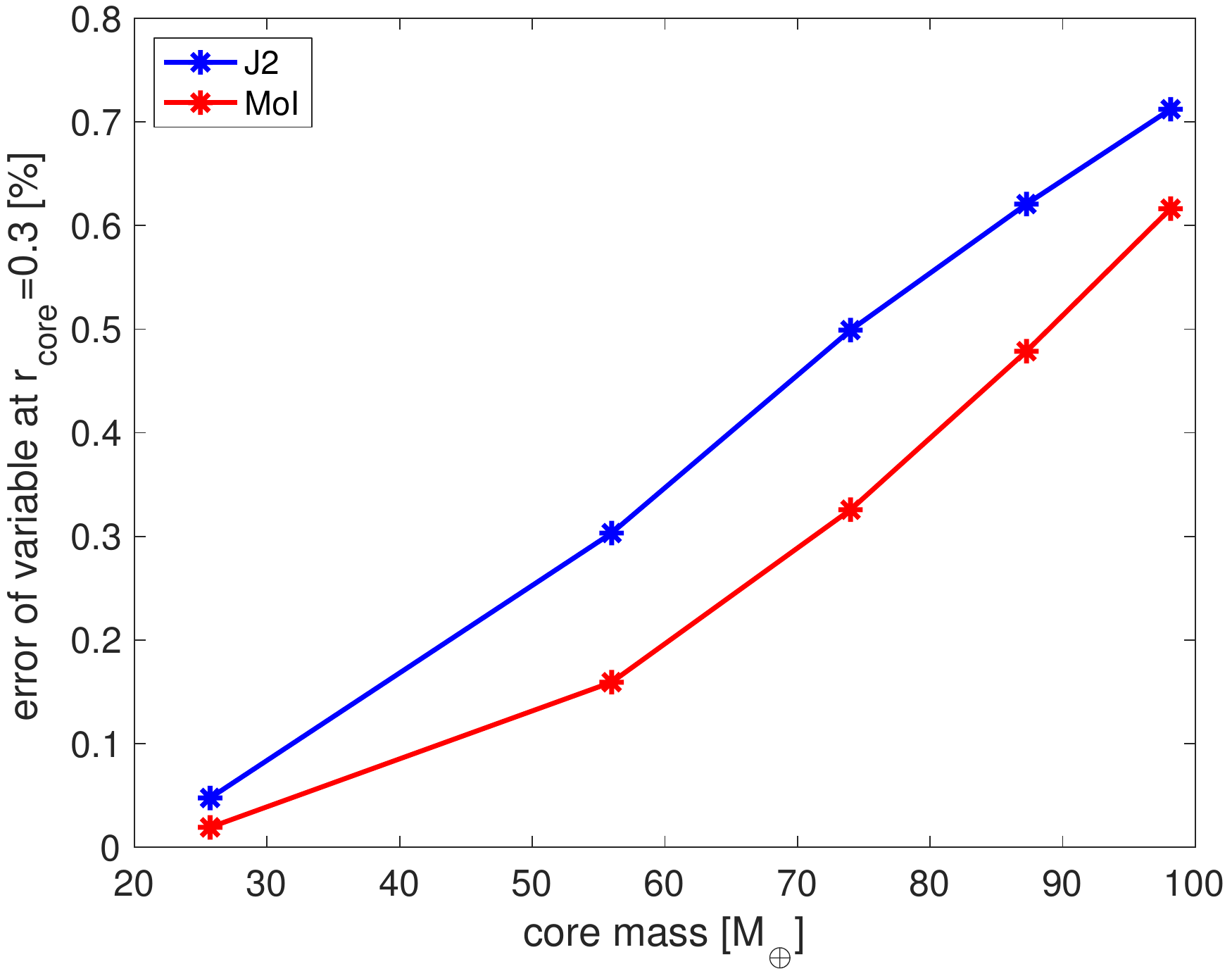}%
    
    \includegraphics[width = 0.5\textwidth]{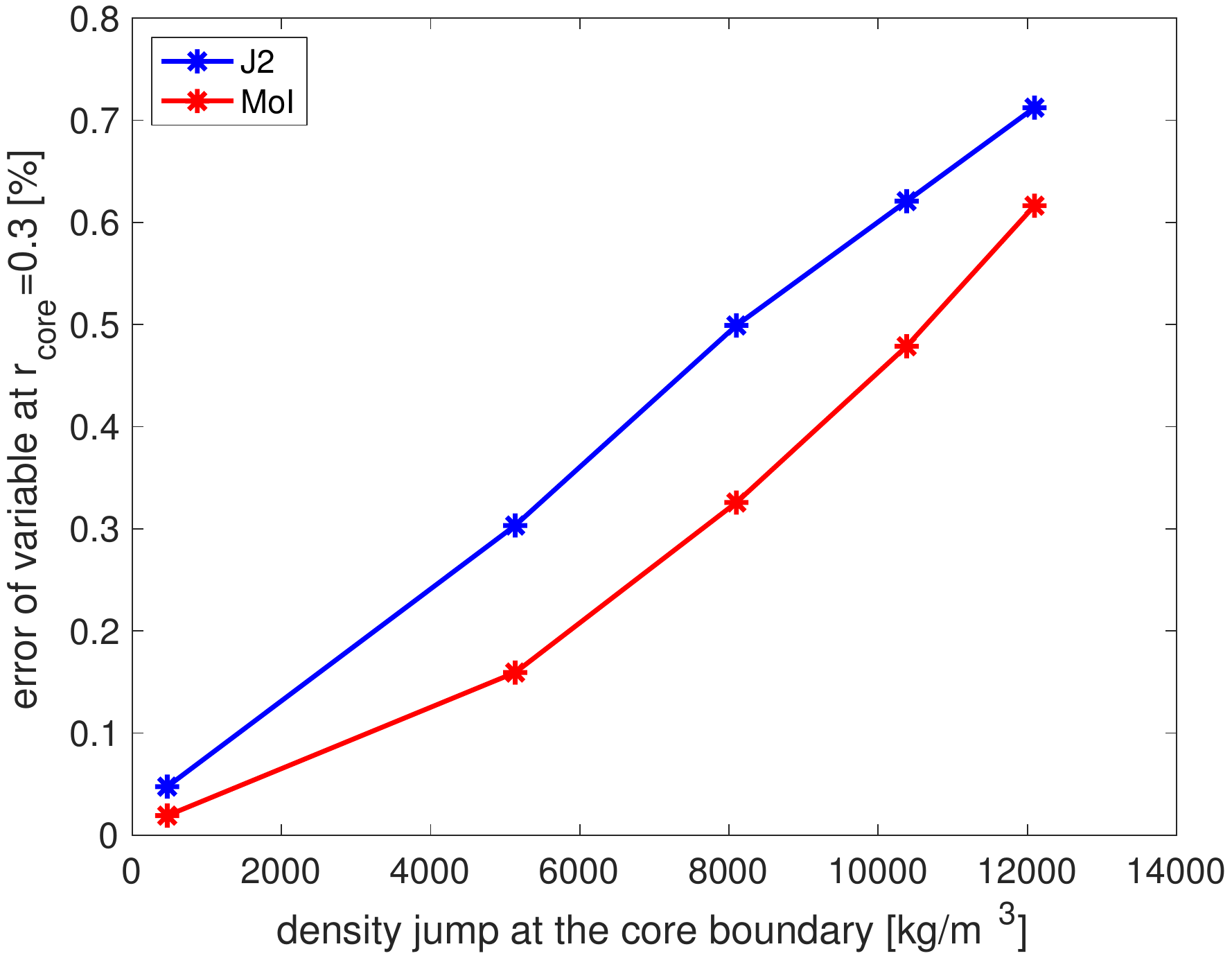}
    \caption{Inferred error in the MoI and $J_2$ depending on either the core mass (upper panel) or the density jump at the core envelope boundary (lower panel) at a fixed core radius of $r\sub{core}=0.3$. We note that in both panels the inferred errors in $J_2$ and the MoI as well as the slopes of the curves are different. This is because the MoI value contains additional information on the density profile that is stored in the higher order $J_{n}$ beyond $J_2$}. The inferred errors in $J_2$ and the MoI are smaller for low mass cores and for (diluted) cores with rather smooth core-envelope transitions. Therefore especially massive and distinct CDC have to be replaced by a PC.
    \label{fig:error_vs_mcore}
\end{figure}

We next investigate how $J_2$ and the MoI are affected by the core mass and the magnitude of the density discontinuity at the core-envelope boundary. 
For this analysis we fix the core radius arbitrarily at $r\sub{core}=0.3$ and show for all five models the inferred errors in $J_2$ and the MoI.
Figure \ref{fig:error_vs_mcore} shows the error of $J_{2}$ (blue dots) and the MoI (red dots) on the y-axis, depending on either the core mass (upper panel) or the magnitude of the density discontinuity at the core-envelope boundary (lower panel). 

First, we observe that $J_2$ and the MoI are not identical. Neither the points overlap nor the slope does agree. This is expected since the MoI, unlike $J_2$, also contains the information of the perturbed higher order $J$-values.
Second, the inferred error of a low-mass core (or a core with a smooth core-envelope transition) is small. However, this error increases for heavy core masses and distinct density jumps at the core-boundary. This leads us to the expected conclusion that especially the most massive CDC and/or the ones with a very distinct density jump at its core-envelope boundary have to be replaced by a PC. Further investigations of this topic are desirable and we hope to address them in future research. 

\section{Resolution Dependent Solutions} \label{section:resoltution_dependence}
The computed planetary shape depends on the resolution used (i.e., number of equipotential levels) and the layer's radial distribution throughout the planet.
As a result, the resolution and distribution used affects the inferred gravitational moments and the MoI. 
Here we test the resolution-dependence of calculated $J_2$, $J_4$ and the MoI by evaluating density profiles of \textit{good results} for various numbers of levels. To diminish potential effects of the spline interpolation (described in section \ref{sec:methods}) on the results, the shape function is evaluated on each equipotential level. 
Table \ref{tab:resolution} summarizes the results using an example. The upper (lower) part shows the calculated gravitational coefficients $J_2$ and $J_4$ and the MoI depending on the number of levels evaluated by ToF (CMS). \\
Independent of the calculation method, the values of $J_2$, $J_4$ and the MoI change significantly for the various numbers of evaluated levels. Nevertheless a convergence is observable for high precision models that use increasing numbers of layers. 
Therefore, it depends on the resolution whether a given density profile represents a planet's gravity field or not. Accordingly, a low-resolution model converges to a different density profile with a different MoI value than a high-resolution solution. This finding is of some importance, as it first limits the ability of compare seemingly similar published results if they are based on different resolutions. Second, a consensus about a minimal resolution has to be reached. For ToF fourth-order we suggest for future studies to evaluate a minimal level number of 2048. For higher resolution, relative changes in $J_2$, $J_4$ and the MoI are in the order of $10^{-4}$ (with respect to the 8192-level result). For CMS no convergence to the method's precision of $10^{-5}$ is found within the tested resolutions.
However, to achieve a precision in the order of $10^{-4}$, 4096 levels are necessary. These recommendations will set studies on the same basis and allows to compare nominal results between them. It is true that more sophisticated schemes can be used to distribute a fixed number of levels along the planet's radius, rather than making them equally spaced. Such schemes can sometimes accelerate convergence to a desired precision level, but at the cost of making it difficult to compare different models to each other. It is also true that different density distributions need a different number of evaluated layers to converge. Therefore it is urgently necessary for each study to test and validate the convergence of their solutions. 
\begin{table*}[htb]
\centering
 \begin{tabular}{c|c c c c c}
 \hline
ToF & 512 & 1024 & 2048 & 4096 & 8192 \\
 \hline

$J_2$ & 0.0147501&	0.0147132&	0.0146991&	0.0146965&	0.0146950 \\
$-J_4$ &0.0005892&	0.0005871&	0.0005868&	0.0005868&	0.0005868\\
MoI	&0.2638440&	0.2638130&	0.2638255&	0.2638653&	0.2638841 \\

\\
\hline
CMS & 512 & 1024 & 2048 & 4096 & 8192 \\
 \hline
$J_2$ & 0.0145916	& 0.0146768	& 0.0146851	& 0.0146886 & 0.01469231\\
$-J_4$ & 0.0005787	& 0.0005847 & 0.0005856 & 0.0005860 & 0.0005863\\
MoI & 0.2630828	& 0.2637095	& 0.2637627	& 0.2637846 & 0.2638103\\
\\
\hline
 \end{tabular}
 \caption{$J_2$, $J_4$ and the MoI depending on different model resolutions (number of equipotential levels). The evaluated internal structure is fixed for this study and based on a \textit{good result}. The evaluation of the $J$-values and the MoI is done with both ToF (upper part) and CMS (lower part).
 }
\label{tab:resolution}
\end{table*}

\end{document}